\documentclass{article}

\PassOptionsToPackage{numbers, compress}{natbib}

\usepackage[final]{neurips_data_2024}

\usepackage[utf8]{inputenc} 
\usepackage[T1]{fontenc}    
\usepackage{hyperref}       
\usepackage{url}            
\usepackage{booktabs}       
\usepackage{amsfonts}       
\usepackage{nicefrac}       
\usepackage{microtype}      
\usepackage[table]{xcolor}         

\usepackage{amsmath}
\usepackage{amssymb}
\usepackage{mathtools}
\usepackage{amsthm}
\usepackage{multirow}
\usepackage{wrapfig}
\usepackage[capitalize,noabbrev]{cleveref}
\usepackage{graphicx}
\usepackage{subfigure} 
\usepackage{float}
\usepackage{algorithm} 
\usepackage{algorithmic}
\usepackage{textcomp} 
\usepackage{listings} 
\usepackage{enumitem}
\usepackage{bbding}
\usepackage{tipa}

\usepackage{array}     

\makeatletter
\renewcommand{\@fnsymbol}[1]{%
  \ifcase#1\or
  \dagger\or 
  *\or       
  \ddagger\or
  \mathsection\or
  \mathparagraph\or
  \|\or
  **\or
  \dagger\dagger\or
  \ddagger\ddagger\else\@ctrerr\fi}
\makeatother

\title{GTSinger: A Global Multi-Technique Singing Corpus with Realistic Music Scores for All Singing Tasks}

\author{%
\normalsize
Yu Zhang\thanks{Equal contribution}\quad
Changhao Pan\footnotemark[1]\quad
Wenxiang Guo\footnotemark[1]\quad
Ruiqi Li\quad
Zhiyuan Zhu\quad
Jialei Wang\quad\\
\normalsize
\textbf{Wenhao Xu} \quad
\textbf{Jingyu Lu}\quad
\textbf{Zhiqing Hong}\quad
\textbf{Chuxin Wang}\quad
\textbf{LiChao Zhang}\quad
\textbf{Jinzheng He}\quad\\
\normalsize
\textbf{Ziyue Jiang}\quad
\textbf{Yuxin Chen}\quad
\textbf{Chen Yang}\quad
\textbf{Jiecheng Zhou}\quad
\textbf{Xinyu Cheng}\quad
\textbf{Zhou Zhao}\thanks{Corresponding Author}\\
 Zhejiang University\\
\normalsize
\texttt{\{yuzhang34,panch,guowx314,zhaozhou\}@zju.edu.cn}\vspace{-1em}
}

\begin{document}

\maketitle

\begin{abstract}

The scarcity of high-quality and multi-task singing datasets significantly hinders the development of diverse controllable and personalized singing tasks, as existing singing datasets suffer from low quality, limited diversity of languages and singers, absence of multi-technique information and realistic music scores, and poor task suitability.
To tackle these problems, we present \textbf{GTSinger}, a large \textbf{G}lobal, multi-\textbf{T}echnique, free-to-use, high-quality singing corpus with realistic music scores, designed for all singing tasks, along with its benchmarks.
Particularly,
(1) we collect 80.59 hours of high-quality singing voices, forming the largest recorded singing dataset;
(2) 20 professional singers across nine widely spoken languages offer diverse timbres and styles;
(3) we provide controlled comparison and phoneme-level annotations of six commonly used singing techniques, helping technique modeling and control;
(4) GTSinger offers realistic music scores, assisting real-world musical composition;
(5) singing voices are accompanied by manual phoneme-to-audio alignments, global style labels, and 16.16 hours of paired speech for various singing tasks.
Moreover, to facilitate the use of GTSinger, we conduct four benchmark experiments: technique-controllable singing voice synthesis, technique recognition, style transfer, and speech-to-singing conversion.
The demos can be found at \url{http://aaronz345.github.io/GTSingerDemo}.
We provide the dataset and the code for processing data and conducting benchmarks at \url{https://huggingface.co/datasets/AaronZ345/GTSinger} and \url{https://github.com/AaronZ345/GTSinger}.

\end{abstract}

\section{Introduction}
\label{sec: intro}

Traditional singing tasks, typically singing voice synthesis (SVS) \cite{kim2023muse,zhang2022wesinger}, aim to generate high-quality singing voices using lyrics and musical notations, attracting broad interest in the industry and academic communities.
As deep learning technology advances, there is a growing demand for more controllable and personalized singing experiences.
This burgeoning demand has catalyzed the emergence of various new singing tasks like technique-controllable SVS, technique recognition, style transfer, and speech-to-singing (STS) conversion \cite{agarwal2022leveraging,li2023alignsts}. 
These tasks have been progressively developed and applied in real life, like short videos and professional composition \cite{zhang2024stylesinger}.

Despite the significant progress made in multiple singing tasks, the scarcity of publicly available high-quality and multi-task singing datasets has become a major bottleneck in their development due to the high cost of recording songs and manual annotations.
The primary limitations of existing open-source singing datasets are as follows:
1) The \textbf{low quality} \cite{ren2020deepsinger} may lead to singing models producing off-pitch, unpleasant, or noisy results.
2) A limited variety in \textbf{languages} \cite{wilkins2018vocalset} and \textbf{singers} \cite{wang2022opencpop} restricts personalized singing models to learn diverse timbres and styles.
3) The absence of the controlled comparison and annotations for multiple \textbf{singing techniques} (like falsetto) \cite{choi2020children}, constrains the technique modeling and control for singing models.
4) The lack of \textbf{realistic music scores} \cite{huang2021multi} hinders human composers from using singing models in real-world musical composition.
5) Poor \textbf{task suitability} \cite{zhang2022m4singer} forces multiple emerging singing tasks to customize new datasets with high cost.

\begin{figure}[t]
\centering
\includegraphics[width=\linewidth]{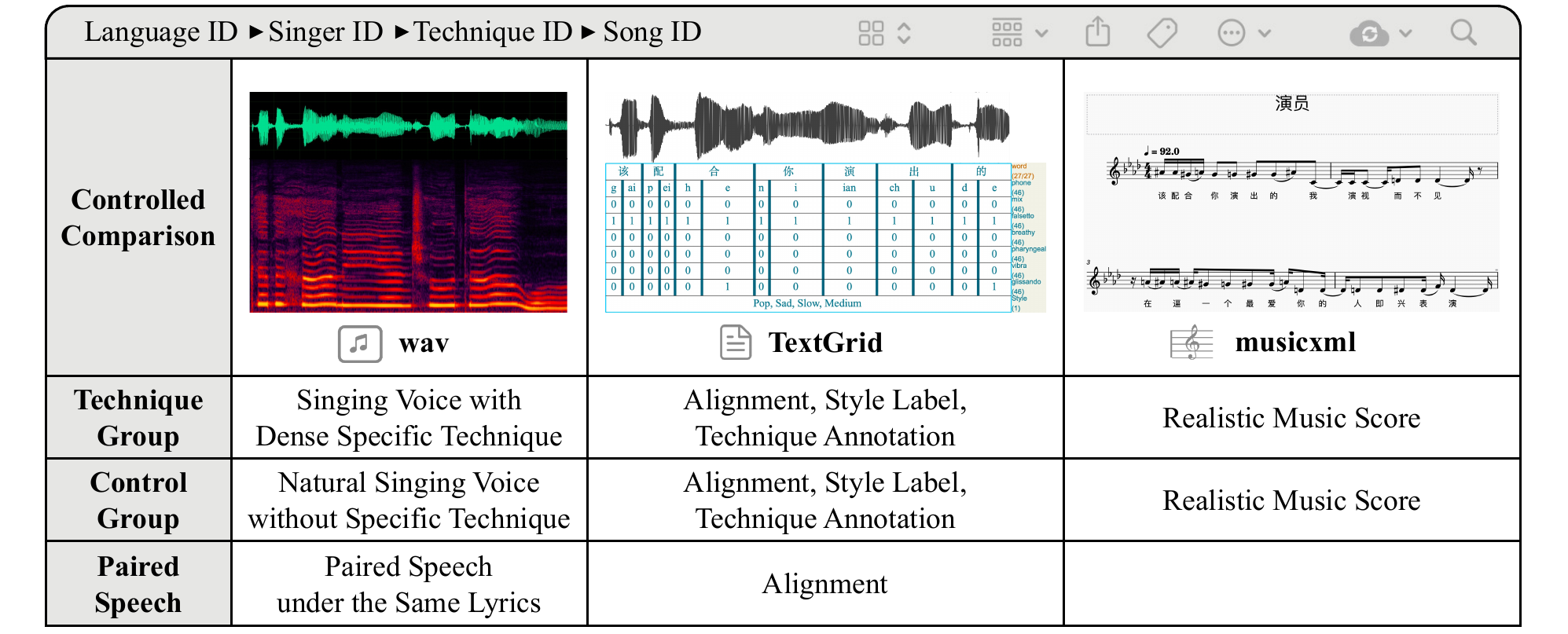}
\vskip-0.1em
\caption{
The composition of each song in GTSinger. 
There are 1,366 songs recorded by 20 singers using six singing techniques across nine languages.
Each song contains a technique group and a control group for the controlled comparison, along with a paired speech for STS tasks.
Alignments, style labels, technique annotations, and realistic music scores are manually created for each group.
}
\label{fig: com}
\end{figure}

To address these challenges, we introduce \textbf{GTSinger}, a large \textbf{G}lobal, multi-\textbf{T}echnique, free-to-use, high-quality singing corpus with realistic music scores, designed for all singing tasks. 
Our dataset contains 80.59 hours of high-quality singing voices without accompaniment, delivered by 20 professional singers covering nine widely spoken languages, including \textbf{Chinese, English, Japanese, Korean, Russian, Spanish, French, German, and Italian}.
Moreover, GTSinger integrates phoneme-level annotations for six commonly used singing techniques, namely \textbf{mixed voice, falsetto, breathy, pharyngeal, vibrato, and glissando}. 
As shown in Figure \ref{fig: com}, each song includes a control group for the natural singing voice and a technique group that intensively applies specific techniques under the same lyrics. 
Additionally, GTSinger furnishes realistic music scores for real-world musical composition.
We also incorporate manual phoneme-to-audio alignments, global style labels (including singing method, emotion, range, and pace), and 16.16 hours of paired speech for various singing tasks. 
Overall, GTSinger boasts several advantages for multiple singing tasks over existing singing datasets:
\begin{itemize}[leftmargin=*]
\item \textbf{80.59 hours of singing voices} in GTSinger are recorded in professional studios by skilled singers, ensuring \textbf{high quality and clarity}, forming the largest recorded singing dataset.
\item Contributed by \textbf{20 singers} across \textbf{nine widely spoken languages} and all four vocal ranges, GTSinger enables zero-shot SVS and style transfer models to learn diverse timbres and styles.
\item GTSinger provides \textbf{controlled comparison} and \textbf{phoneme-level annotations} of \textbf{six singing techniques} for songs, thereby facilitating singing technique modeling, recognition, and control.
\item Unlike fine-grained music scores, GTSinger features \textbf{realistic music scores} with regular note duration, assisting singing models in learning and adapting to real-world musical composition.
\item The dataset includes \textbf{manual phoneme-to-audio alignments}, \textbf{global style labels}, and \textbf{16.16 hours of paired speech}, ensuring comprehensive annotations and broad task suitability.
\end{itemize}

The rest of the paper is organized as follows. 
In Section \ref{sec: rel}, we briefly review and compare with current singing datasets. 
In Section \ref{sec: data}, we provide the construction details and data statistics of GTSinger.
In Section \ref{sec: ben}, to demonstrate the use and validate the quality of GTSinger, 
we conduct extensive experiments and establish benchmarks for four different singing tasks, including \textbf{technique-controllable singing voice synthesis, technique recognition, style transfer, and STS conversion}, employing recently published state-of-the-art methods for each task. 
In Section \ref{sec: con}, we make the conclusion and discuss some potential risks along with the limitations of GTSinger.

\section{Related Work}
\label{sec: rel}

The advancement of deep learning has enabled singing models, like singing voice synthesis (SVS) models, to achieve remarkably high-quality vocal results \cite{kim2023muse,zhang2022wesinger,ren2020deepsinger,zhang2022visinger,liu2022diffsinger,he2023rmssinger}. 
Fueled by the growing demand for controllable and personalized singing experiences, diverse new singing tasks have emerged, like technique-controllable SVS, technique recognition, style transfer, and speech-to-singing (STS) conversion \cite{jayashankar2023self,takahashi2022robust,agarwal2022leveraging,li2023alignsts,liu2022learning,zhang2024stylesinger}. 
Unlike traditional datasets \cite{ren2020deepsinger,duan2013nus,tamaru2020jvs} designed for a singular task, a high-quality and multi-task singing dataset has higher demands. 
A high-quality dataset not only requires recordings by professional singers with manual alignments but also needs to include multiple singers and languages to broaden timbres and styles. 
Furthermore, a multi-task dataset also needs to feature controlled comparisons and phoneme-level annotations of singing techniques for technique modeling, realistic music scores for real-world musical composition, global style labels for global control, and paired speech for STS tasks. 
These requirements significantly elevate the recording and annotation costs, explaining the scarcity of high-quality and multi-task datasets.

\begin{table}[ht]
\centering
\small
\caption{The information table of existing open-source singing datasets.
Align and RMS mean manual phoneme-to-audio alignment and realistic music scores.
Style denotes global style labels.
}
\scalebox{0.9}{
\begin{tabular}{l|c|c|cc|cccc|c}
\toprule
 \multirow{2}{*}{\bfseries{Corpus}} & \multirow{2}{*}{\textbf{Language}} & \multirow{2}{*}{\textbf{Singer}} & \multicolumn{2}{c|}{\textbf{Hours}} & \multicolumn{4}{c|}{\textbf{Manual Annotations}} & \multirow{2}{*}{\parbox{1.45cm}{\centering \textbf{Controlled} \\ \textbf{Comparison}}} \\ 
 & & & Singing & Speech & Align & RMS & Tech & Style & \\
\midrule
 VocalSet \cite{wilkins2018vocalset} & 1 & 20 & 10.1 & 0 & \XSolidBrush & \XSolidBrush & \XSolidBrush & \XSolidBrush & \Checkmark \\
  CSD \cite{choi2020children} & 2 & 1 & 4.86 & 0 & \XSolidBrush & \XSolidBrush & \XSolidBrush & \XSolidBrush  & \XSolidBrush \\
 KVT \cite{kim2020semantic} & 1 & 114 & 18.85 & 0 & \XSolidBrush & \XSolidBrush & \XSolidBrush & \Checkmark & \XSolidBrush \\
  PopBuTFy \cite{liu2022learning} & 2 & 34 & 50.8 & 0 & \XSolidBrush & \XSolidBrush & \XSolidBrush & \XSolidBrush & \XSolidBrush \\
  OpenSinger \cite{huang2021multi} & 1 & 66 & 50 & 0 & \XSolidBrush & \XSolidBrush & \XSolidBrush  & \XSolidBrush & \XSolidBrush \\
 NHSS \cite{sharma2021nhss} & 1 & 10 & 4.75 & 2.25 & \XSolidBrush & \XSolidBrush & \XSolidBrush  & \XSolidBrush & \XSolidBrush \\
 Tohoku Kiritan \cite{ogawa2021tohoku} & 1 & 1 & 1 & 0 & \Checkmark & \XSolidBrush & \XSolidBrush  & \XSolidBrush  & \XSolidBrush \\
  OpenCpop \cite{wang2022opencpop} & 1 & 1 & 5.25 & 0 & \Checkmark & \XSolidBrush & \XSolidBrush & \XSolidBrush & \XSolidBrush \\
  M4Singer \cite{zhang2022m4singer} & 1 & 20 & 29.77 & 0 & \Checkmark & \XSolidBrush & \XSolidBrush & \XSolidBrush  & \XSolidBrush \\
\midrule
 {\bfseries{GTSinger (Ours)}} & \textbf{9} & \textbf{20} & \textbf{80.59} & \textbf{16.16} & \Checkmark & \Checkmark & \Checkmark & \Checkmark & \Checkmark \\
\bottomrule
\end{tabular}}
\label{tab: data}
\end{table}

As delineated in Table \ref{tab: data}, several datasets endeavor to mitigate specific challenges to cater to designated tasks. 
VocalSet \cite{wilkins2018vocalset} provides the controlled comparison of various singing techniques, albeit constrained by its reliance solely on a range of vowels rather than linguistic content in singing voices.
CSD \cite{choi2020children} features recordings in both English and Korean, yet a limited number of vocalists constrains its diversity. 
KVT \cite{kim2020semantic} annotates some types of global style labels in K-pop songs but uses existing songs and does not separate vocals from accompaniments.
PopBuTFy \cite{liu2022learning} provides singing voices in both English and Chinese, but without annotations.
OpenSinger \cite{huang2021multi} encompasses a substantial volume of vocal recordings across numerous singers, yet it does not contain any annotation. 
NHSS \cite{sharma2021nhss} introduces paired speech for STS tasks but falls short in providing manual phoneme-level alignments and other annotations. 
Tohoku Kiritan \cite{ogawa2021tohoku} provides manual alignments but is limited by its small scale.
Opencpop \cite{wang2022opencpop} and M4Singer \cite{zhang2022m4singer} mark significant advancements with their manual alignments and music scores. 
However, they only provide fine-grained music scores, which disrupt the regularity of note duration and thus, hinder the application to real-world musical composition. 
Moreover, they lack other annotations and paired speech for more tasks. 
In this paper, we construct a large multi-lingual, multi-singer, free-to-use, high-quality singing corpus with controlled comparison and phoneme-level annotations of multiple techniques, along with manual phoneme-to-audio alignments, realistic music scores, global style labels, and paired speech.
We seek to comprehensively address the limitations in previous singing datasets and cater to all current singing tasks.

\section{Dataset Description}
\label{sec: data}

\begin{figure}[ht]
\centering
\includegraphics[width=\linewidth]{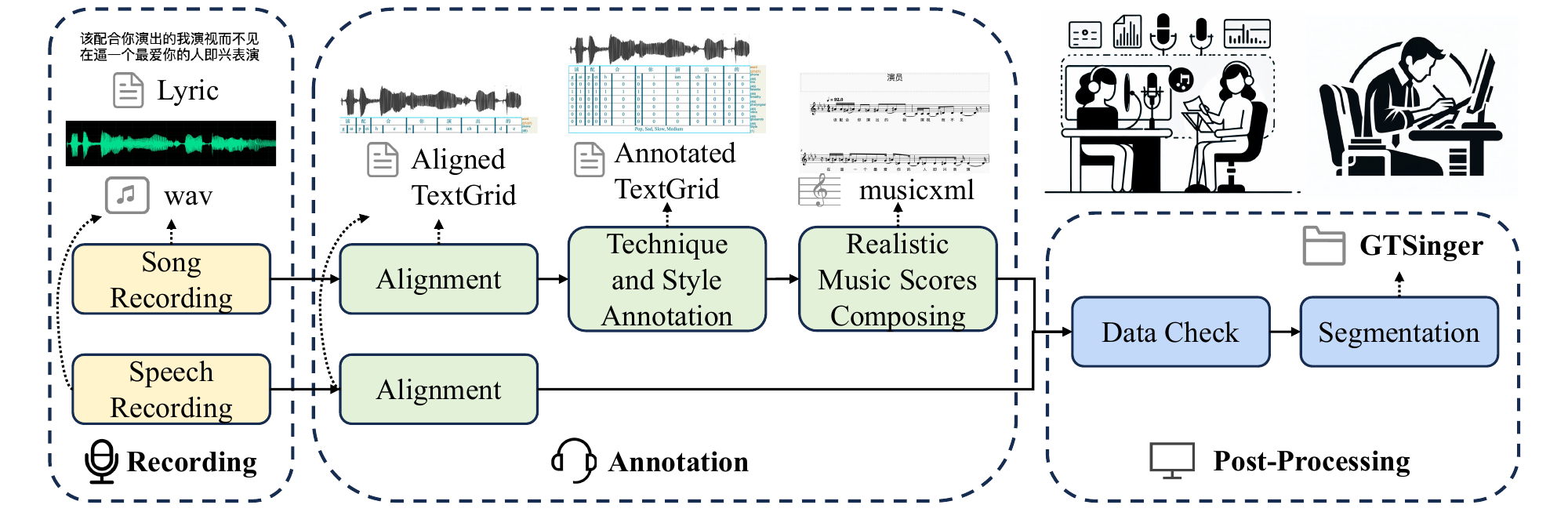}
\vskip-0.1em
\caption{
The pipeline of data collection of GTSinger.
Human double-checks exist in each process.
}
\label{fig: pip}
\end{figure}

In this section, we formally introduce GTSinger, a large global, multi-technique, free-to-use, high-quality singing corpus with realistic music scores, which aims to support all current singing tasks and can be used under license CC BY-NC-SA 4.0.
Figure \ref{fig: pip} depicts the pipeline of the creation of GTSinger, with detailed explanations of each step provided in the following subsections. 
Then, we present the necessary dataset statistics to enhance the understanding of our GTSinger.
At last, we also provide the instructions to use our dataset and codes.

\subsection{Songs and Singers}

To construct GTSinger, we first select nine widely spoken languages: Chinese, English, Japanese, Korean, Russian, Spanish, French, German, and Italian. 
Then, we also choose six commonly used singing techniques: mixed voice, falsetto, breathy, pharyngeal, vibrato, and glissando. 
After rigorous auditions, we select 20 professional singers, covering all four vocal ranges (alto, soprano, tenor, bass), and each singer is proficient in all six techniques and one of the widely spoken languages.
Before recording, all singers agree to make their vocal performances open-source for academic research.
We carefully select songs based on the representativeness of each language, the vocal range of each singer, and the suitability of singing each technique densely.
As shown in Table \ref{tab: song}, all singers are listed and anonymized by their languages along with vocal ranges.

\begin{table}[ht]
\centering
\small
\caption{The information table of languages, singers, techniques, and duration. 
Singing hours for technique ID count time for singing voices in both control groups and technique groups.}
\scalebox{0.88}{
\begin{tabular}{l|c|cc|ccccc}
\toprule
\multirow{3}{*}{\textbf{Language ID}} & \multirow{3}{*}{\textbf{Singer ID}} & \multicolumn{2}{c|}{\bfseries{Total Hours}} & \multicolumn{5}{c}{\bfseries{Singing Hours of Technique ID}} \\ 
& & \multirow{2}{*}{Singing} & \multirow{2}{*}{Speech} & \multirow{2}{*}{\parbox{1.6cm}{\centering Mixed Voice \\ and Falsetto}} & \multirow{2}{*}{Breathy} & \multirow{2}{*}{Pharyngeal} & \multirow{2}{*}{Vibrato} & \multirow{2}{*}{Glissando} \\ 
& & & & & & & & \\
\midrule
\multirow{2}{*}{\textbf{Chinese (ZH)}} & {ZH-Tenor-1} & 8.45 & 1.82 & 3.6 & 1.26 & 1.18 & 1.18 & 1.23 \\
& {ZH-Alto-1} & 8.14 & 1.49 & 3.7 & 1.13 & 1.06 & 1.13 & 1.12 \\
\midrule
\multirow{3}{*}{\textbf{English (EN)}} & {EN-Tenor-1} & 4.76 & 0.87 & 2.06 & 0.69 & 0.65 & 0.7 & 0.66 \\
& {EN-Alto-1} & 3.47 & 0.67 & 1.6 & 0.52 & 0.51 & 0.28 & 0.56 \\
& {EN-Alto-2} & 4.9 & 1.04 & 2.05 & 0.74 & 0.67 & 0.73 & 0.71 \\
\midrule
\multirow{2}{*}{\textbf{Japanese (JA)}} & {JA-Tenor-1} & 2.13 & 0.29 & 1.01 & 0.33 & 0.34 & 0.15 & 0.3 \\
& {JA-Soprano-1} & 4.32 & 0.87 & 2.24 & 0.56 & 0.41 & 0.53 & 0.58 \\
\midrule
\multirow{3}{*}{\textbf{Korean (KO)}} & {KO-Tenor-1} & 4.61 & 1.32 & 1.19 & 0.87 & 0.88 & 0.83 & 0.84 \\
& {KO-Soprano-1} & 0.95 & 0.24 & 0.19 & 0.16 & 0.2 & 0.21 & 0.19 \\
& {KO-Soprano-2} & 2.72 & 0.61 & 1.12 & 0.37 & 0.42 & 0.42 & 0.39 \\
\midrule
\textbf{Russian (RU)} & {RU-Alto-1} & 4.32 & 0.76 & 1.81 & 0.63 & 0.55 & 0.7 & 0.63 \\
\midrule
\multirow{2}{*}{\textbf{Spanish (ES)}} & {ES-Bass-1} & 4.45 & 0.9 & 2.01 & 0.61 & 0.61 & 0.61 & 0.61 \\
& {ES-Soprano-1} & 3.48 & 0.82 & 1.4 & 0.59 & 0.4 & 0.53 & 0.56 \\
\midrule
\multirow{2}{*}{\textbf{French (FR)}} & {FR-Tenor-1} & 4.58 & 0.58 & 1.27 & 0.9 & 0.84 & 0.66 & 0.91 \\
& {FR-Soprano-1} & 3.96 & 0.59 & 1.75 & 0.58 & 0.58 & 0.57 & 0.48 \\
\midrule
\multirow{2}{*}{\textbf{German (DE)}} & {DE-Tenor-1} & 4.54 & 0.9 & 2.19 & 0.56 & 0.59 & 0.59 & 0.61 \\
& {DE-Soprano-1} & 4.54 & 0.82 & 1.9 & 0.64 & 0.63 & 0.67 & 0.7 \\
\midrule
\multirow{3}{*}{\textbf{Italian (IT)}} & {IT-Bass-1} & 3.21 & 0.82 & 0.86 & 0.76 & 0.17 & 0.68 & 0.74 \\
& {IT-Bass-2} & 1.61 & 0.4 & 0.32 & 0.32 & 0.3 & 0.33 & 0.34 \\
& {IT-Soprano-1} & 1.45 & 0.35 & 0.98 & 0.11 & 0.1 & 0.05 & 0.21 \\ 
\midrule
\textbf{All} & \textbf{All} & \bf{80.59} & \bf{16.16} & \bf{33.25} & \bf{12.33} & \bf{11.09} & \bf{11.55} & \bf{12.37} \\
\bottomrule
\end{tabular}}
\label{tab: song}
\end{table}

\subsection{Recording}

Singers perform a multitude of songs, each selected to highlight a specific singing technique (like falsetto). 
For each song, they maintain a consistent rhythm, lyrics, and key, recording twice: once densely applying the specific technique (technique group) and once for the natural singing voice without the specific technique(control group). 
We especially manage falsetto and mixed voice techniques due to their correlations. 
They form a distinct group, recording a natural singing voice (control group), and two technique groups, for both falsetto and mixed voice. 
Furthermore, each song includes an additional spoken lyric sentence recorded by the same singer, providing paired speech for STS tasks. 
All recordings are carried out in a professional studio, with singers listening to the song's accompaniment through headphones, ensuring clean vocal tracks devoid of accompaniment yet preserving rhythm and timing. 
Each audio is recorded at a 48kHz sampling rate with 24 bits in WAV format, ensuring high-quality data for further statistics and research.
Table \ref{tab: song} presents the duration of 1,366 final recorded songs.
For more details, please refer to Appendix \ref{app: re}.

\subsection{Annotation}


\textbf{Alignment:}
We initially use the Montreal Forced Aligner (MFA) \cite{mcauliffe2017montreal} for a coarse alignment of the original lyrics and audio and store the results in TextGrid format. 
Chinese phonemes are extracted using pypinyin, English phonemes follow the ARPA standard, Italian phonemes follow the Epitran standard, and others follow the MFA standard.
These are the most effective and suitable phoneme standards for these languages.
Next, annotators with a musical background use Praat \cite{boersma2001praat} to correct the rough annotation results, focusing on the following areas:
(1) Boundary correction: Annotators correct the boundaries of words and phonemes by listening to the audio and observing the mel-spectrogram, which forms the bulk of this step.
(2) Word and phoneme correction: In cases of missing or incorrect lyrics, annotators are required to correct the words and corresponding phonemes based on their auditory perception. 
This is because singers may mispronounce words, or there may be homophones in Chinese that cause phoneme errors.
(3) Unvoiced labeling: The unvoiced region, including breathing and silent sections, is marked by annotators who identify the boundaries respectively.
In this step, we perform alignment for both the singing voice and paired speech.

\textbf{Technique and Style Annotation:}
Following the alignment process, we instruct our annotators to perform phoneme-level annotations of six singing techniques on the TextGrid, including mixed voice, falsetto, breathy, pharyngeal, vibrato, and glissando.
Annotators continue to use Praat \cite{boersma2001praat} for annotations based on their auditory perception, indicating the presence or absence of each technique for every phoneme. 
Notably, in technique groups, singing voices employ densely specific techniques but not exclusively, as other techniques may also be used.
In control groups, specific techniques are excluded but other techniques can be present as singers are asked to sing naturally. 
Next, annotators also label the singing method (pop and bel canto), emotion (happy and sad), pace (slow, moderate, and fast), and range (low, medium, and high) as global style labels for each group. 

\begin{figure}[ht]
\centering
\includegraphics[width=\linewidth]{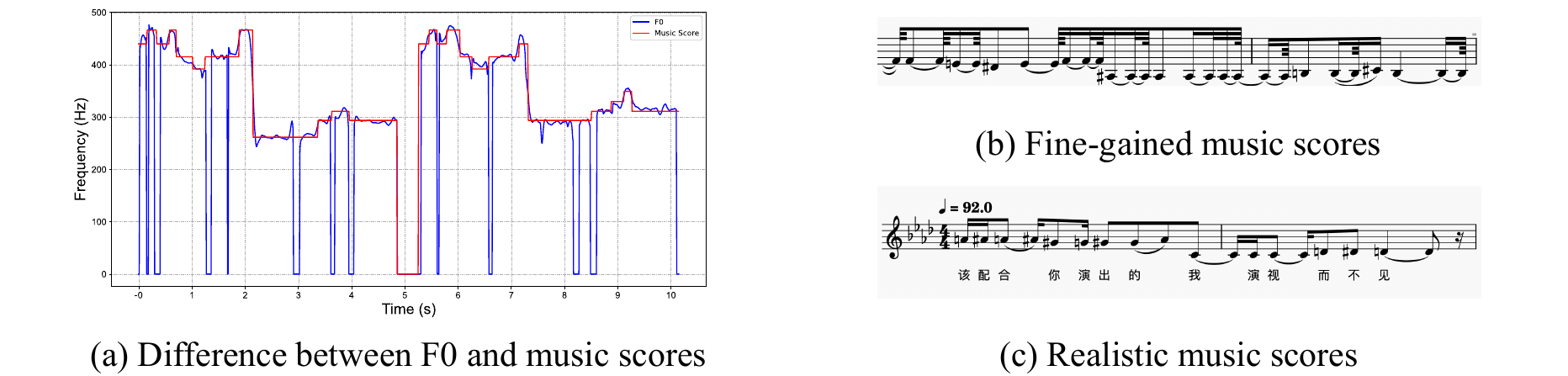}
\vskip-0.1em
\caption{
The comparison between F0, fine-grained music scores, and realistic music scores.
Score pitches are converted to frequencies and are very different from F0.
Fine-grained music scores disrupt the regularity of note duration, resulting in fragmented notes that are unsuitable for composing.
}
\label{fig: rms}
\end{figure}

\textbf{Realistic Music Score Composing:}
The difference in F0, fine-grained music scores, and realistic music scores are depicted in Figure \ref{fig: rms}.
To compose realistic music scores, we initially employ RMVPE \cite{wei2023rmvpe} to extract F0 for each singing voice. 
Then, we use ROSVOT \cite{li2024robust} to derive the MIDI form of the music scores. 
The MIDI is obtained by referring to the F0 curve for determining the note pitch and duration.
Subsequently, we engage music experts to listen to the recorded songs, refer to original accompaniments, and carry out the following steps:
1) Determine the actual tempo, clef, and key.
2) Adjust the music scores to match the true note pitch.
3) Modify the note duration following regular realistic music score rules.
4) Annotate the note type to be rest, lyric, or slur.
The outcome is realistic music scores in the muxicxml format.
More annotation details can be found in Appendix \ref{app: an}.

\subsection{Post-Processing}

\textbf{Data Check:}
For each language with fully annotated data, we employ an additional music expert proficient in that language to randomly inspect 25\% of the annotations. 
Their primary tasks include:
(1) Checking alignment, including word and phoneme boundaries, incorrect characters, polyphonic phonemes in Chinese data, and annotations of unvoiced sections.
(2) Examining technique and style annotations, focusing on annotations of techniques outside the specific group.
(3) Reviewing realistic music scores, paying attention to key, tempo, and clef, and correcting note pitch and duration.

\textbf{Segmentation:}
After completing the data annotation and inspection, we segment the audio into smaller fragments to facilitate training for singing tasks. 
For the same song, the control group, technique group, and paired speech are synchronously segmented into sentence-level segments, with their alignments, annotations, and scores correspondingly segmented. 
By leveraging the manual alignment results, we set a threshold for the unvoiced region and established maximum and minimum lengths for the voiced region as the conditions for performing the segmentation process.
As shown in Figure \ref{fig: stat} (a), we ensure more than 95\% sentences are between 5 and 20 seconds in duration. 
Finally, we get 29,261 singing utterances and 12,373 speech utterances.

\subsection{Statistics:}

\begin{figure}[ht]
\centering
\includegraphics[width=\linewidth]{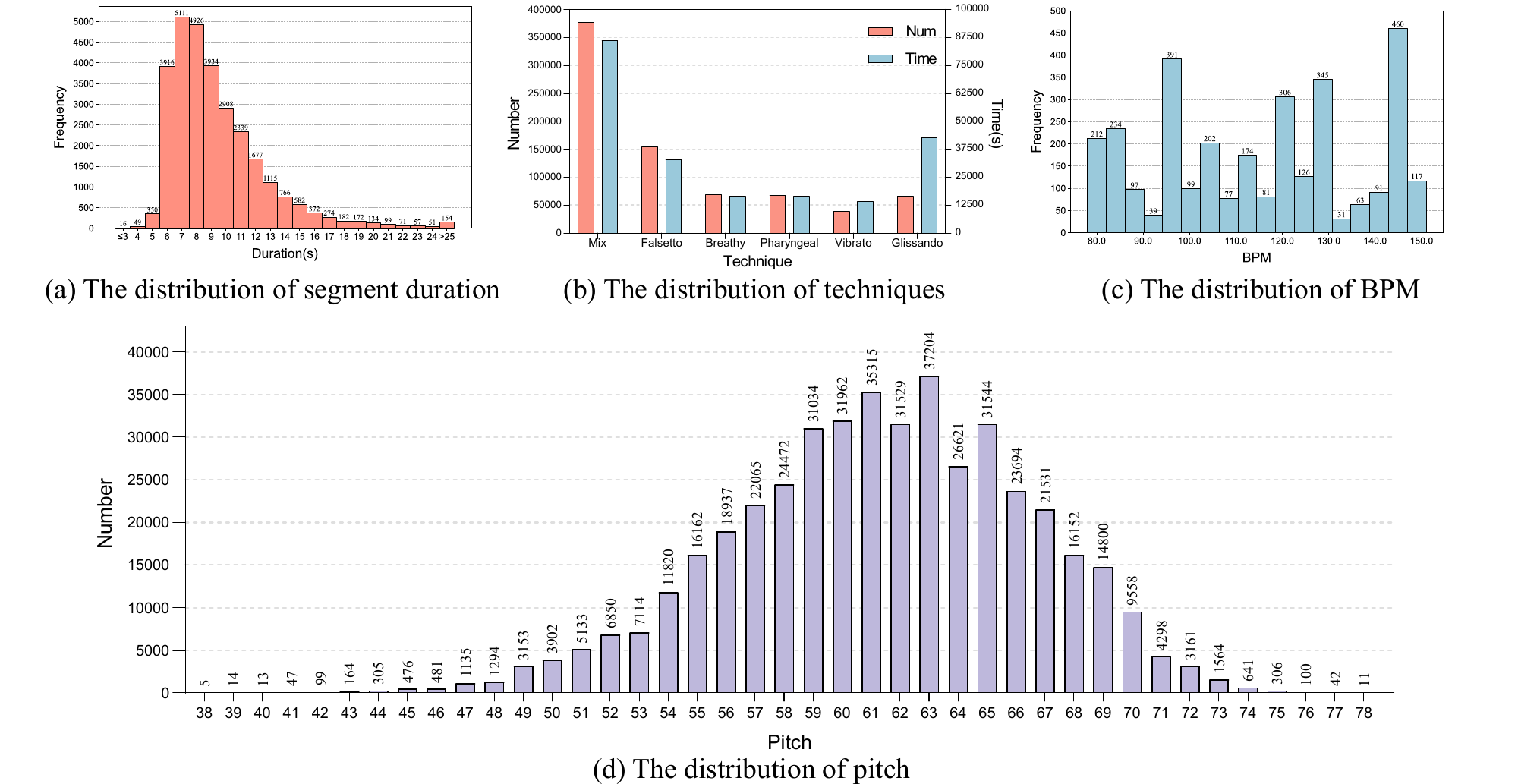}
\vskip-0.1em
\caption{
The statistical distribution of segment duration, techniques, BPM, and pitch.
}
\label{fig: stat}
\end{figure}

To illustrate the diverse range of recorded singing voices, Figure \ref{fig: stat} (b) presents the distribution of six singing techniques. 
The prevalence of mixed voice can be attributed to its widespread use in both pop and bel canto singing methods. 
Conversely, vibrato is the least frequent because it can only be used on sustained notes, resulting in a very low distribution density.
Figure \ref{fig: stat} (c) presents the distribution of beats per minute (BPM) across 3,145 groups (each song contains technique and control groups). 
The majority of groups fall within the 80 to 150 BPM range, encompassing the primary range for both pop and bel canto singing methods.
Figure \ref{fig: stat} (d) illustrates that note pitches of realistic music scores are primarily distributed between MIDI note numbers 50 (D3, 146.83 Hz) and 70 (B4, 494 Hz), covering four vocal ranges.
This broad spectrum of languages, singers, techniques, BPM, and pitch suggests the potential of models trained on GTSinger to effectively handle a diverse range of singing styles. 
Refer to Appendix \ref{app: stat} for further detailed statistics.

\subsection{Instructions for Use}

The dataset can be freely downloaded at \url{https://huggingface.co/datasets/AaronZ345/GTSinger} and noncommercially used under license CC BY-NC-SA 4.0. 
Users can define additional tasks under this license. 
We also provide the code for processing data at \url{https://github.com/AaronZ345/GTSinger}.
For suggestions, please contact us via email. 
Regular updates will be provided on our GitHub repository.
\section{Benchmarks}
\label{sec: ben}

In this section, to assess the quality and versatility of GTSinger, we conduct a comprehensive evaluation across four singing tasks: technique-controllable SVS, technique recognition, style transfer, and speech-to-singing (STS) conversion.
We evaluate GTSinger with recently published state-of-the-art methods, utilizing four NVIDIA 2080Ti GPUs for all experimental protocols. 
Hifi-GAN \cite{kong2020hifi} is employed as the vocoder for audio synthesis. 
We conduct the MOS (mean opinion score), FFE, MCD, and Cos for evaluation of these tasks on the test set.
For more details about the evaluation, please refer to Appendix \ref{app: sub} and \ref{app: obj}.
For more detailed results, please refer to Appendix \ref{app: exp}.

\subsection{Technique-Controllable Singing Voice Synthesis}
\label{sec: svs}

Previous SVS models are limited by datasets and cannot achieve technique-controllable SVS. 
Technique-controllable SVS can provide more controllable and personalized singing experiences for real-world applications, allowing even those who cannot sing to customize a variety of professional techniques.
We randomly choose five songs from each singer, totaling 100 songs as the test set, with the remainder as the training set.
We conduct experiments on the following systems:
(1) GT: The ground truth singing voice;
(2) GT (vocoder): The audio generated by the pre-trained HiFi-GAN;
(3) DiffSinger \cite{liu2022diffsinger}: A popular SVS system based on the diffusion model;
(4) RMSSinger \cite{he2023rmssinger}: An outstanding SVS system using the diffusion model to predict F0.
(5) StyleSinger \cite{zhang2024stylesinger}: the current state-of-the-art SVS system using the residual style encoder to model both global and detailed styles.
We enhance both models with phoneme-level technique embedding to achieve technique control.

We employ both objective and subjective evaluation metrics for evaluation.
For subjective evaluation, MOS-Q indicates the quality, naturalness, and clarity of the synthesized audio, while MOS-C reflects the expressiveness and accuracy of technique controllability. 
Both metrics are rated on a scale from 1 to 5 and reported with 95\% confidence intervals.
For objective evaluation, we use F0 Frame Error (FFE) to measure the accuracy of F0 and UV prediction, and Mean Cepstral Distortion (MCD) for audio quality measurement.
We conduct both parallel and non-parallel experiments according to the target technique sequence.
In the parallel experiments, we use the GT technique sequence as the target. 
In the non-parallel experiments, six techniques are randomly yet appropriately assigned to each target phoneme (zero, one or more techniques on each phoneme).

\begin{table}[ht]
\centering
\small
\caption{
Technique-controllable SVS performance in both parallel and non-parallel experiments. 
We use FFE, MCD, MOS-Q, and MOS-C for comparisons.
}
\scalebox{1}{
\begin{tabular}{l|cccc|cc}
\toprule
\multirow{2}{*}{\bfseries{Method}} & \multicolumn{4}{c|}{\bfseries{Parallel}} & \multicolumn{2}{c}{\bfseries{Non-Parallel}}\\
& {FFE $\downarrow$}  & {MCD $\downarrow$} & {MOS-Q $\uparrow$} & {MOS-C $\uparrow$} & {MOS-Q $\uparrow$} & {MOS-C $\uparrow$} \\
\midrule
GT & - & - & 4.54 $\pm$ 0.06 & - & -& -\\
GT (vocoder) & 0.05 & 1.33 & 4.21 $\pm$ 0.07 & 4.42 $\pm$ 0.03 &- & -\\
\midrule
DiffSinger \cite{liu2022diffsinger} & 0.29 & 3.58 & 3.81 $\pm$ 0.06 & 3.83 $\pm$ 0.07 & 3.77 $\pm$ 0.05 & 3.78 $\pm$ 0.07 \\
RMSSinger \cite{he2023rmssinger} & 0.27 & 3.43 & 3.94 $\pm$ 0.07 & 3.95 $\pm$ 0.05 & 3.86 $\pm$ 0.06 & 3.89 $\pm$ 0.06 \\
StyleSinger \cite{zhang2024stylesinger} & 0.25 & 3.27 & 4.01 $\pm$ 0.09 & 4.15 $\pm$ 0.06 & 3.95 $\pm$ 0.08 & 4.10 $\pm$ 0.05 \\
\bottomrule
\end{tabular}}
\label{tab: svs}
\end{table}

As shown in Table \ref{tab: svs}, we can observe the following: 
(1) Leveraging the diffusion decoder, DiffSinger achieves reasonable sound quality (MOS-Q). 
However, it struggles to model and control techniques effectively (MOS-C). 
Additionally, the high FFE and VDE indicate that the generated styles deviate significantly from the actual curve.
(2) By using a diffusion model to model F0, RMSSinger significantly improves its ability to handle techniques (MOS-C) and achieves better synthesis quality (MOS-Q). 
This highlights the impact of pitch modeling on technique representation.
(3) StyleSinger integrates multi-level style information from the reference mel-spectrograms, and renders techniques more naturally and expressively, outperforming all other baseline models across all metrics.
This demonstrates the complexity involved in modeling techniques across different singing styles.
However, these results indicate that there is still a significant gap between the model's performance and GT (vocoder), highlighting ample room for improvement in the technique controllability of singing tasks. 
Future work can explore using more advanced generation models to incorporate realistic music scores and leverage phoneme-level technique information for better F0 and mel-spectrogram generation, as well as higher technique controllability. 
For more detailed and visualized results about technique-controllable SVS, please refer to Appendix \ref{app: tc}.

\subsection{Technique Recognition}
\label{sec: tr}

Technique recognition aims to predict the techniques in unseen audio samples, facilitating the augmentation of existing singing datasets with technique annotations and aiding the real-world learning of singing techniques.
We design a technique recognition model based on ROSVOT \cite{li2024robust} and change the loss to the cross entropy loss of each technique label.
The inputs of the technique recognition model include the mel-spectrogram, pitch, and phoneme boundaries, with the output being the predicted probabilities of six techniques in each phoneme.
We conduct both overall and cross-lingual experiments to evaluate our model's performance and the annotation quality of GTSinger. 
We categorize the languages into two groups: Asian (Chinese, Japanese, and Korean) and European (Italian, Spanish, English, French, German, and Russian).
In overall experiments, We reuse the rule in Section \ref{sec: svs} to split training and test sets.
In cross-lingual experiments, models are trained on one group of languages (like Asian) and tested on the other language group (like European) to assess their generalization capabilities.
For evaluation, we provide F1 and Accuracy.

\begin{table}[ht]
\centering
\small
\caption{
F1 and Accuracy of each technique in overall and cross-lingual technique recognition. 
}
\scalebox{1}{
\begin{tabular}{l|c|cccccc}
\toprule
\multirow{2}{*}{\bfseries{Experiment}} & \multirow{2}{*}{\bfseries{Metric}} & \multicolumn{6}{c}{\textbf{Technique Recognition Accuracy}}\\
& &{mixed voice} & {falsetto} &{breathy} & {pharyngeal} & {vibrato} &{glissando} \\
\midrule
\multirow{2}{*}{\textbf{Overall}} & F1 & 0.78 & 0.96 & 0.99 & 0.85 & 0.70 & 0.70 \\
& Accuracy & 0.78 & 0.84 & 0.78 & 0.80 & 0.89 & 0.85 \\
\midrule
\multirow{2}{*}{\textbf{Cross-Lingual}} & F1 & 0.72 & 0.94 & 0.96 & 0.84 & 0.66 & 0.64 \\
& Accuracy & 0.75 & 0.79 & 0.72 & 0.77 & 0.84 & 0.78 \\
\bottomrule
\end{tabular}}
\label{tab: tr}
\end{table}

As shown in Table \ref{tab: tr}, our model demonstrates good F1 and Accuracy of all six techniques in both overall and cross-lingual experiments, highlighting the quality of our technique recognition model, as well as the merit of designing controlled comparison and phoneme-level annotation of six techniques in GTSinger. 
However, it is evident that the overall performance still surpasses that of cross-lingual cases. 
This indicates ample room for improvement in the technique recognition model, suggesting the potential to enhance generalization, thus handling out-of-domain technique recognition better.
For more details about the model and results, please refer to Appendix \ref{app: tr}.

\subsection{Style Transfer}
\label{sec: st}

Style transfer aims to generate high-quality singing voices with the timbre and styles (like singing methods, rhythm, techniques, and pronunciation) of the reference audio. 
This technology can be applied in the dubbing of entertainment short videos, offering personalized experiences.
We reuse the rule in Section \ref{sec: svs} to split training and test sets.
For baseline models, we reuse StyleSinger, as it is the first singing style transfer model, and enrich RMSSinger with additional singer and emotion embedding for conducting style transfer like previous works \cite{zhang2024stylesinger}.
For subjective evaluation, we use MOS-Q for synthesis quality and MOS-S for singer similarity in terms of timbre and styles. 
For objective evaluation, we employ FFE to measure pitch accuracy, MCD to assess synthesis quality, and Cos for evaluation of singer similarity.
We conduct both parallel and cross-lingual style transfer experiments.
For parallel experiments, we use another singing voice by the same singer as the reference audio.
For cross-lingual experiments, we also split languages into Asian and European groups like Section \ref{sec: tr}.
Then we randomly select reference audio from one language group (like Asian), transferring singing styles to target lyrics in the other language group (like European).

\begin{table}[ht]
\centering
\small
\caption{
Style Transfer performance in both parallel and cross-lingual experiments. 
We use FFE, MCD, Cos, MOS-Q, and MOS-C for comparisons.
}
\scalebox{1}{
\begin{tabular}{l|ccccc|cc}
\toprule
\multirow{2}{*}{\bfseries{Method}} & \multicolumn{5}{c|}{\bfseries{Parallel}} & \multicolumn{2}{c}{\bfseries{Cross-Lingual}}\\
& {FFE $\downarrow$}  & {MCD $\downarrow$} & {Cos $\uparrow$} & {MOS-Q $\uparrow$} & {MOS-S $\uparrow$} & {MOS-Q $\uparrow$} & {MOS-S $\uparrow$} \\
\midrule
GT & - & - & - & 4.53 $\pm$ 0.03 & - & -& -\\
GT (vocoder) & 0.05 & 1.34 & 0.96 & 4.18 $\pm$ 0.04 & 4.26 $\pm$ 0.03 &- & -\\
\midrule
RMSSinger  & 0.31 & 3.47 & 0.88 & 3.70 $\pm$ 0.04 & 3.79 $\pm$ 0.06 & 3.66 $\pm$ 0.04 & 3.76 $\pm$ 0.08\\
StyleSinger  & 0.26 & 3.29 & 0.93 & 3.95 $\pm$ 0.06 & 4.01 $\pm$ 0.05 & 3.89 $\pm$ 0.07 & 3.92 $\pm$ 0.09 \\
\bottomrule
\end{tabular}}
\label{tab: st}
\end{table}

As shown in Table \ref{tab: st}, we can observe that StyleSinger achieves impressive results in both synthesized quality (MOS-Q) and singer similarity (MOS-S), which suggests that GTSinger's extensive style collection facilitates modeling and transfer, enabling the model to achieve high performance.
Additionally, StyleSinger performs well in cross-lingual tasks, showcasing its ability to sing any music scores and lyrics, regardless of the singer's identity. 
However, there is still ample room for improvement in the cross-lingual style transfer performance. 
Future research can explore specialized models for handling singing style transfer tasks involving significant style differences.
For more detailed and visualized results about style transfer, please refer to Appendix \ref{app: st}.

\subsection{Speech-to-Singing Conversion}
\label{sec: sts}

Speech-to-singing (STS) conversion aims to transform speech into the corresponding singing voice preserving the timbre and phoneme information. 
STS can be applied to automatic music production or personalized entertainment.
We randomly select five songs (including paired speech) from each singer, totaling 100 pairs as the test set, and others as the training set.
Besides GT and GT (vocoder), we use AlignSTS \cite{li2023alignsts} as a baseline model, and design another model based on StyleSinger, which inputs paired speech as the reference audio to conduct STS conversion. 
StyleSinger uses realistic musical scores (RMS), which are more practical, while AlignSTS requires GT singing F0 as input.
We reuse evaluation metrics in Section \ref{sec: st} for evaluating synthesized quality and singer similarity.

\begin{table}[ht]
\centering
\small
\caption{
Speech-to-singing performance in FFE, MCD, Cos, MOS-Q, and MOS-S metrics.
}
\scalebox{1}{
\begin{tabular}{l|ccc|ccc}
\toprule
{\bfseries{Method}} & {FFE $\downarrow$} & {MCD $\downarrow$} & {Cos $\uparrow$} & {MOS-Q $\uparrow$} & {MOS-S $\uparrow$} \\
\midrule
GT & - & - & - & 4.53 $\pm$ 0.03 & - \\
GT (vocoder) & 0.05 & 1.34 & 0.95 & 4.17 $\pm$ 0.05 & 4.20 $\pm$ 0.04 \\
\midrule
AlignSTS  & 0.35 & 3.52 & 0.85 & 3.68 $\pm$ 0.12 & 3.73 $\pm$ 0.09 \\
StyleSinger & 0.28 & 3.38 & 0.92 & 3.83 $\pm$ 0.09 & 3.88 $\pm$ 0.08\\
\bottomrule
\end{tabular}}
\label{tab: sts}
\end{table}

As shown in Table \ref{tab: sts}, we observe that StyleSinger outperforms AlignSTS in both synthesis quality (MOS-Q) and singer similarity (MOS-S).
This also demonstrates the quality of our annotations of realistic music scores.
Using realistic music scores and speech to synthesize expected singing voices allows for more controllable and personalized singing experiences in real-world applications. 
We can observe the potential for improvement in speech-to-singing performance, indicating that there is ample room for enhancement in pitch modeling based on realistic music scores.
Future work can explore using realistic music scores for better pitch modeling, as well as designing specialized intermediate models to better convert styles between speech and singing styles.
For more detailed and visualized results about the speech-to-singing conversion, please refer to Appendix \ref{app: sts}.
\section{Conclusion and Discussion}
\label{sec: con}

In this paper, we propose a novel dataset GTSinger, a large global, multi-technique, free-to-use, high-quality singing corpus with realistic music scores, designed for all singing tasks, comprehensively addressing the limitations of existing singing datasets. 
Furthermore, we provide the construction process and statistical analysis for GTSinger. 
In addition, we have conducted extensive experiments and established four benchmarks, thereby contributing further to future singing research.

\textbf{Limitations and Future Directions:}
(1) Our dataset currently lacks comprehensive coverage of widely spoken languages, like Arabic, and does not include several commonly used singing techniques, such as vocal fry.
Future efforts will be directed towards expanding the diversity of the singing data.
(2) Although our annotation process is performed by professionals with musical expertise, accurately segmenting phoneme durations within words and identifying subtle singing techniques remains challenging for human ears. 
Future models that better utilize word-level annotations may mitigate some of the errors introduced by manual labeling.
(3) While our dataset addresses various singing tasks, extending its utility to the broader music field may require integration with vocal-to-accompaniment models like SingSong \cite{donahue2023singsong} to assist in generating music that includes vocals.

\textbf{Negative Societal Impact:} 
The presence of sensitive biometric data in our dataset inherently carries potential risks. 
Therefore, We first perform data desensitization and consider using techniques such as vocal watermarking to further protect personal privacy. 

\section*{Acknowledgements}
This work was supported by National Natural Science Foundation of China under Grant No.62222211 and Grant No.62072397.

\bibliographystyle{neurips_2024}
\bibliography{neurips_2024}


\newpage
\appendix
\begin{center}{\bf {\LARGE Appendices} }
\end{center}
\begin{center}{\bf {\Large GTSinger: A Global Multi-Technique Singing Corpus \\ with Realistic Music Scores for All Singing Tasks} \linebreak}
\end{center}

\section{Details of Dataset}
\label{app: stat}

\subsection{Details of Recording}
\label{app: re}

We have recruited 20 professional singers, each proficient in at least one language, and signed formal contracts with them. 
They are hired at a rate of \$300 per hour of audio recording to perform specified language skill songs. 
In total, we spend \$30,000 on the recording process.
Before recording, all singers agree to make their vocal performances open-source for academic research. All recordings are conducted in a professional studio, with singers listening to the song's accompaniment through headphones. 
This ensures clean vocal tracks without accompaniment while preserving rhythm and timing. 
Each audio is recorded at a 48kHz sampling rate with 24-bit depth in WAV format, ensuring high-quality data for further statistical analysis and research.
The 20 singers cover all four vocal ranges: tenor, alto, bass, and soprano. 
They also perform in nine widely spoken languages: Chinese, English, Japanese, Korean, Russian, Spanish, French, German, and Italian.
We require the singers to perform controlled comparison recordings using six singing techniques: mixed voice, falsetto, breathy, pharyngeal, vibrato, and glissando. 
Each technique covers controlled comparisons of multiple songs. 
For each song, they maintain consistent rhythm, lyrics, and key, recording twice: once densely applying the specific technique (technique group) and once using their natural singing voice without the specific technique (control group).
We uniquely manage the falsetto and mixed voice techniques due to their strong correlation, necessitating special contrast for subsequent research. 
These techniques form a distinct group, recording a natural singing version (control group) and two versions for technique groups, for both falsetto and mixed voice. 
The falsetto version can raise the key of the same song to better showcase the falsetto technique while maintaining other consistencies like rhythm.
Furthermore, each song includes an additional spoken lyric sentence recorded by the same singer, providing paired speech for speech-to-singing tasks.

\subsection{Details of Annotation}
\label{app: an}

We hire numerous experts with backgrounds in music and language for our annotation and review process, compensating each at a rate of \$15 per hour. 
In total, we spend \$18,000 on the annotation process. 
Before beginning their tasks, each expert is informed about the use of the annotated data, and they agree to make their annotation results open-source for academic research.
Initially, some experts organize the submitted data from the singers into the required format to facilitate subsequent annotation. 
Then, we use the Montreal Forced Aligner (MFA) \cite{mcauliffe2017montreal} for a coarse alignment of the original lyrics and audio, storing the results in TextGrid format. 
Chinese phonemes are extracted using pypinyin \footnote{https://github.com/mozillazg/python-pinyin}, English phonemes follow the ARPA standard \footnote{https://en.wikipedia.org/wiki/ARPABET}, Italian phonemes follow the Epitran standard \footnote{https://github.com/dmort27/epitran}, while others follow the MFA standard \footnote{https://mfa-models.readthedocs.io/en/latest/dictionary/}. 
We choose these standards because Chinese uses pinyin for pronunciation, ARPA includes English stress patterns, Epitran performs better in Italian phonemes, and the MFA dictionaries better capture the phonetic characteristics of other languages.
Next, annotators use Praat \cite{boersma2001praat} to correct the rough annotation results. 
Following the alignment process, we instruct our annotators to perform phoneme-level annotations of six singing techniques on the TextGrid, including mixed voice, falsetto, breathy, pharyngeal, vibrato, and glissando.
Then, annotators also need to label the singing method (pop and bel canto), emotion (happy and sad), pace (slow, moderate, and fast), and range (low, medium, and high) as global style labels for each group. 
To compose realistic music scores, we initially employ RMVPE \cite{wei2023rmvpe} to extract F0 and ROSVOT\cite{li2024robust} to derive the MIDI form of the scores. 
Subsequently, we engage music experts to listen to the recorded songs, refer to original accompaniments, and annotate realistic music scores in the musicxml format. 
Each step is double-checked by other music experts.
Finally, for each language data, we employ an additional music expert proficient in that language to randomly inspect 25\% of the annotations.

\subsection{Statistics of Global Styles}
\label{app: sty}

\begin{figure}[ht]
\centering
\includegraphics[width=\linewidth]{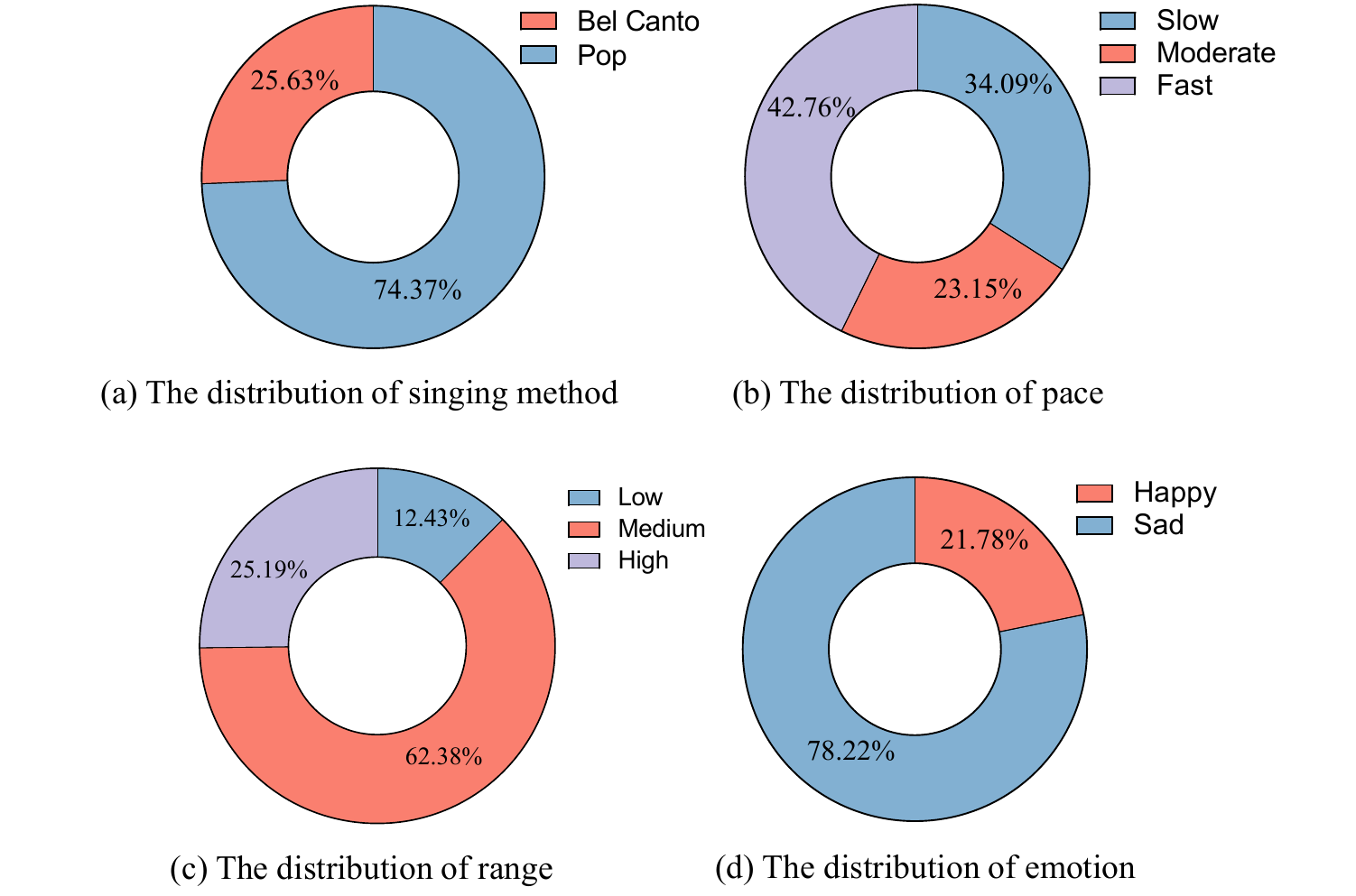}
\vskip
-0.1em
\caption{
The statistical distribution of global style labels.
}
\label{fig: sty}
\end{figure}

We annotate each group (each song comprising control and technique groups) with global style labels provided by music experts. 
These labels include singing method (pop and bel canto), emotion (happy and sad), pace (slow, moderate, and fast), and range (low, medium, and high). 
As shown in Figure \ref{fig: sty}, we have summarized the distribution of these four labels.
It can be observed that GTSinger predominantly features songs with the pop singing method, reaching 74.37\%. 
This is because we selected representative songs from various languages, which are mainly contemporary and widely popular. 
Characteristics of pop songs, such as a higher occurrence of fast pace, a predominant pitch range in the medium category, and a higher frequency of sad emotions, align with our statistical results. 
We can observe that fast pace accounts for 42.67\%, medium range reaches 62.38\%, and sad emotion reaches 78.22\%.
Additionally, the inclusion of diverse types of global styles demonstrates the stylistic variety present in GTSinger.

\subsection{Statistics of Chinese}
\label{app: zh}

\begin{figure}[ht]
\centering
\includegraphics[width=\linewidth]{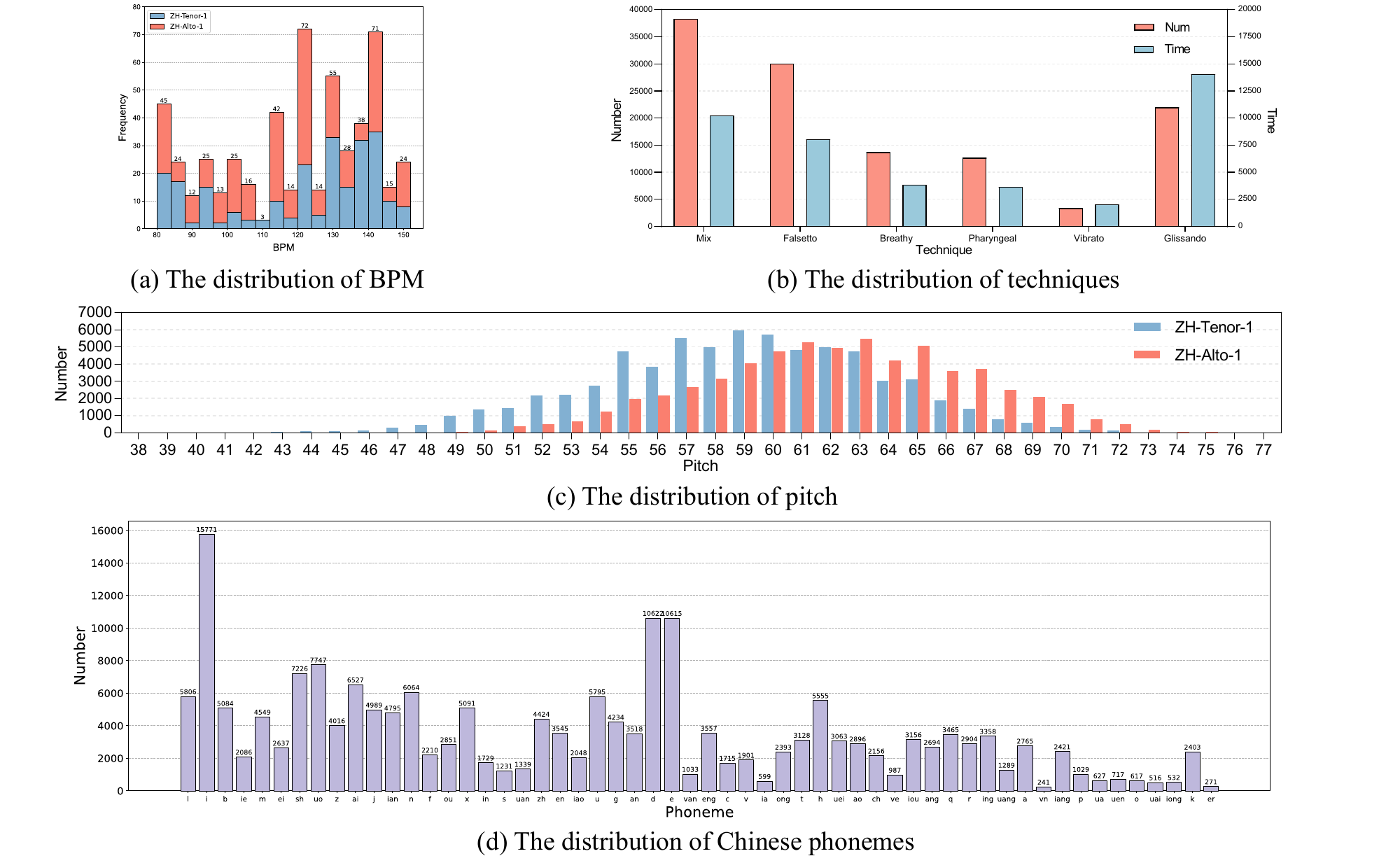}
\vskip-0.1em
\caption{
The statistical distribution of the BPM, techniques, pitch, and phonemes in Chinese.
}
\label{fig: zh}
\end{figure}

There are two Chinese singers, namely ZH-Tenor-1 and ZH-Alto-1. 
As shown in Figure \ref{fig: zh} (a), the overall BPM distribution ranges from 80 to 150, with a dense distribution between 110 and 150 due to the frequent use of their pop singing method, which is typically faster.
As illustrated in Figure \ref{fig: zh} (b), the two singers performed a variety of six techniques, with mixed voice being the most prevalent. 
This predominance is because mixed voice is the most commonly used technique. 
Vibrato is the least frequent because it can only be used on sustained notes, resulting in a very low distribution density.
Figure \ref{fig: zh} (c) shows that the overall note pitch ranges of realistic music scores span from MIDI note number 52 (E3, 164.81 Hz) to 69 (A4, 440 Hz). 
ZH-Tenor-1 primarily ranges from 52 (E3, 164.81 Hz) to 66 (F\#4, 370 Hz), peaking at 59 (B3, 246.94 Hz), and ZH-Alto-1 primarily ranges from 55 (G3, 196 Hz) to 69 (A4, 440 Hz), peaking at 63 (D\#4, 311.13 Hz).
This distribution aligns with their vocal ranges.
As mentioned above, Chinese phonemes are entirely annotated according to pypinyin 
, comprising a total of 56 phonemes. 
This annotation fully adapts to the pronunciation rules of Chinese and represents the most suitable set of Chinese phonemes, as used in previous Chinese singing datasets \cite{zhang2022m4singer}.
Figure \ref{fig: zh} (d) illustrates that the most common phoneme is "i", with 15,771 occurrences, while the least common is "vn", with 241 occurrences. 
The broad distribution of phonemes is highly suitable for singing models.

\subsection{Statistics of English}
\label{app: en}

\begin{figure}[ht]
\centering
\includegraphics[width=\linewidth]{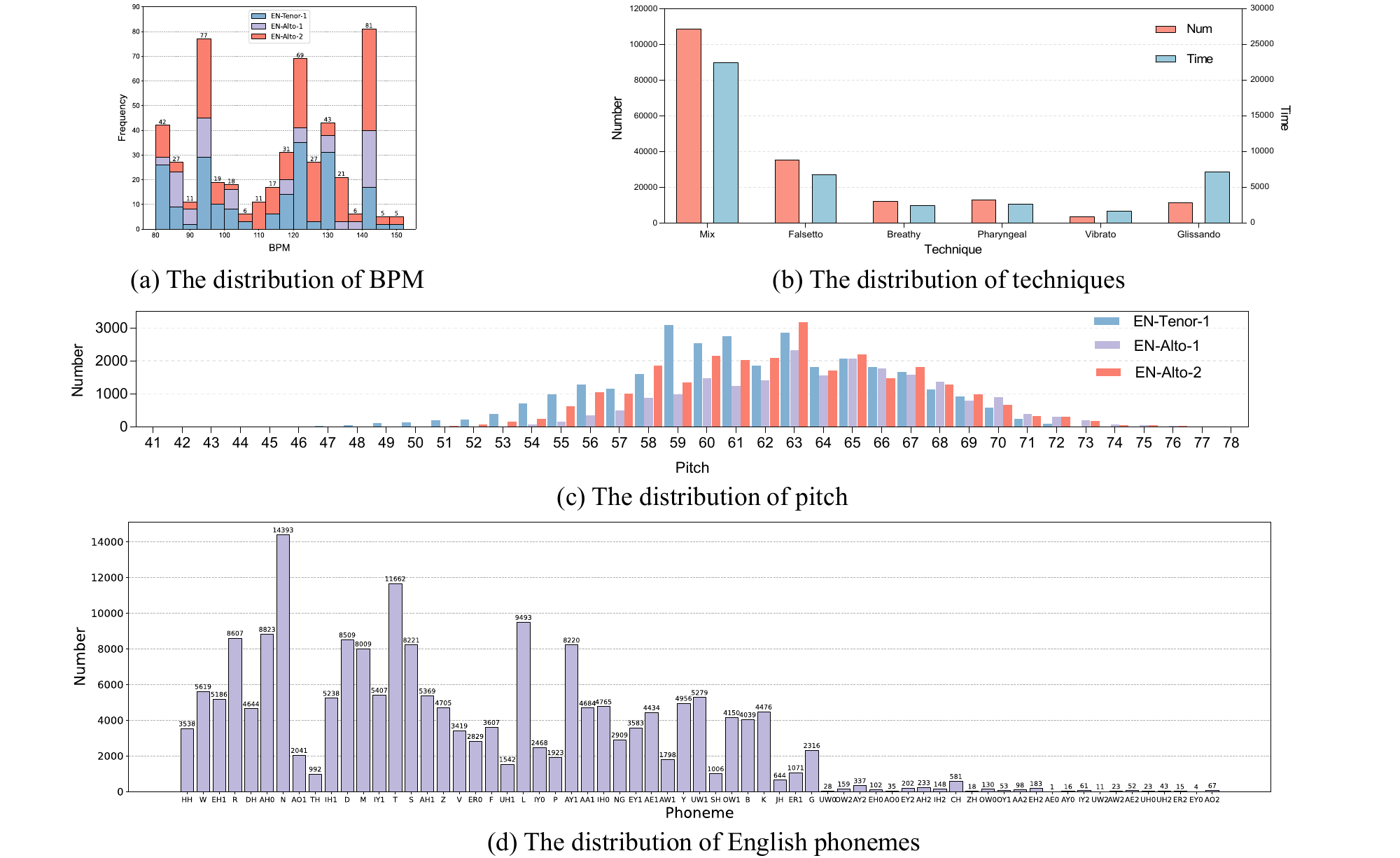}
\vskip-0.1em
\caption{
The statistical distribution of the BPM, techniques, pitch, and phonemes in English.
}
\label{fig: en}
\end{figure}

There are three English singers, namely EN-Tenor-1, EN-Alto-1, and EN-Alto-2. 
As shown in Figure \ref{fig: en} (a), the overall BPM distribution ranges from 80 to 150.
As illustrated in Figure \ref{fig: en} (b), the three singers performed a variety of six techniques. 
Like Chinese, mixed voice is the most prevalent and vibrato is the least used technique.
Figure \ref{fig: en} (c) shows that the overall note pitch ranges of realistic music scores span from MIDI note number 55 (G3, 196 Hz) to 69 (A4, 440 Hz). 
EN-Tenor-1 primarily ranges from 55 (G3, 196 Hz) to 68 (G\#4, 415.30 Hz), peaking at 59 (B3, 246.94 Hz). 
EN-Alto-1 primarily ranges from 59 (B3, 246.94 Hz) to 68 (G\#4, 415.30 Hz), peaking at 63 (D\#4, 311.13 Hz). 
EN-Alto-2 primarily ranges from 57 (A3, 220 Hz) to 69 (A4, 440 Hz), peaking at 63 (D\#4, 311.13 Hz).
This distribution aligns with their vocal ranges.
As mentioned above, English phonemes are entirely annotated according to the ARPA standard.
The ARPA standard includes English stress patterns, therefore, it's appropriate for English phoneme modeling.
Figure \ref{fig: en} (d) illustrates that the most common phoneme is "N", with 14,393 occurrences, while the least common phoneme is "AE0", with 1 occurrence. 
The missing phonemes are "AW0", "OY2", "AA0", and "OY0", which are barely seen in English songs, and the total usage frequency of all words containing them in daily life is less than 0.1\%. 
The same applies to phonemes with an occurrence of less than 10.

\subsection{Statistics of Japanese}
\label{app: ja}

\begin{figure}[ht]
\centering
\includegraphics[width=\linewidth]{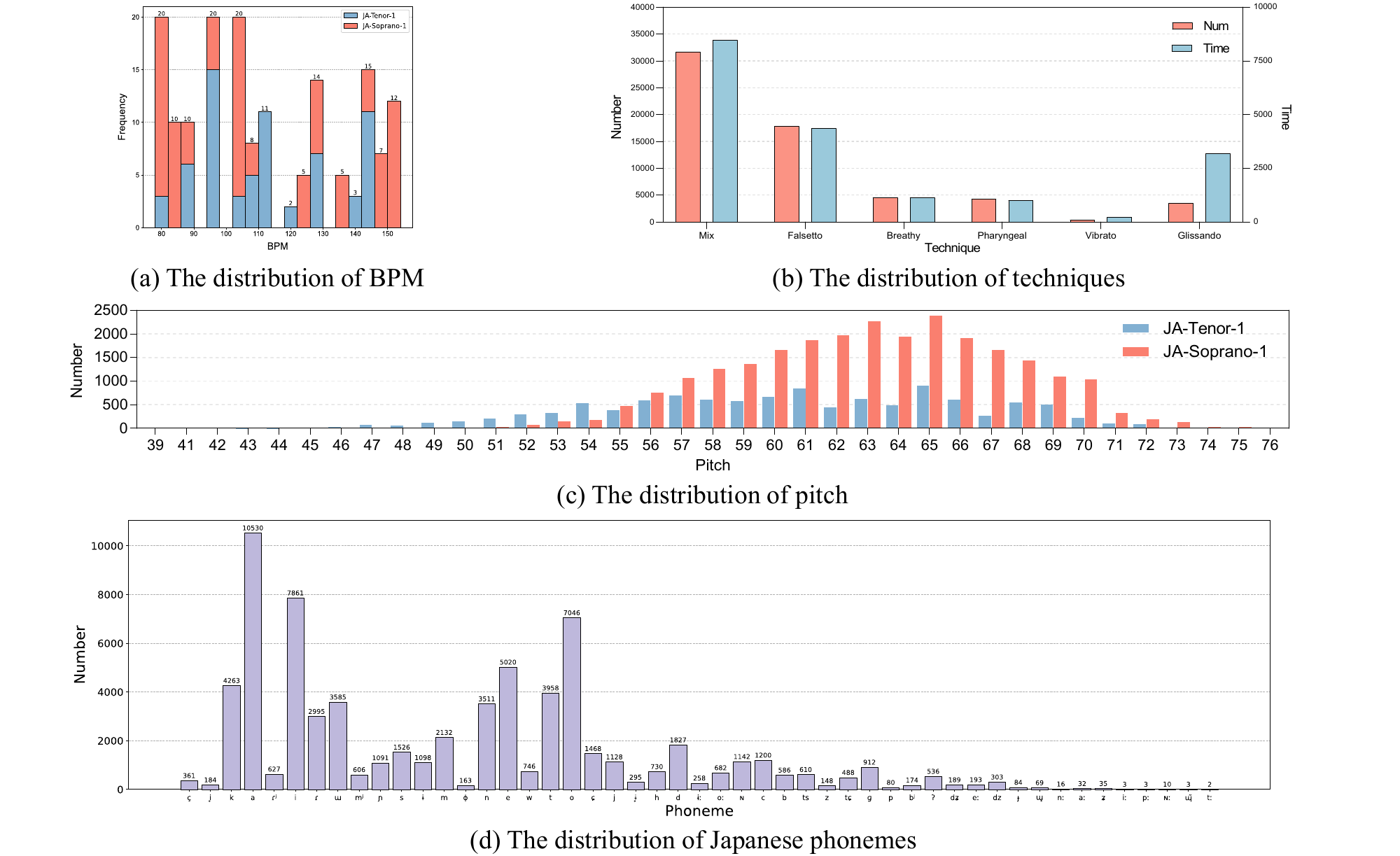}
\vskip-0.1em
\caption{
The statistical distribution of the BPM, techniques, pitch, and phonemes in Japanese.
}
\label{fig: ja}
\end{figure}

There are two Japanese singers, namely JA-Tenor-1 and JA-Soprano-1. 
As shown in Figure \ref{fig: ja} (a), the overall BPM distribution ranges from 80 to 150. 
As illustrated in Figure \ref{fig: ja} (b), the two singers perform a variety of six techniques, with mixed voice being the most prevalent and vibrato being the least used technique. 
Figure \ref{fig: ja} (c) shows that the overall note pitch ranges of realistic music scores span from MIDI note number 55 (G3, 196 Hz) to 70 (A\#4, 466.16 Hz). 
JA-Tenor-1 primarily ranges from 54 (F\#3, 185 Hz) to 69 (A4, 440 Hz), peaking at 65 (F4, 349.23 Hz). 
JA-Soprano-1 primarily ranges from 57 (A3, 220 Hz) to 70 (A\#4, 466.16 Hz), peaking at 65 (F4, 349.23 Hz).
Since the popular songs in Japanese typically have a relatively high range, we choose two singers whose vocal range is high. 
And this distribution aligns with their vocal ranges.
As mentioned above, Japanese phonemes are entirely annotated according to the MFA standard.
Figure \ref{fig: ja} (d) illustrates that the most common phoneme is "a", with 10,530 occurrences, while the least common phoneme is "\textipa{t\textlengthmark}", with 2 occurrences.
Phonemes with an occurrence of less than 10 are rarely found in Japanese songs. 
Additionally, the total usage frequency of all words containing these phonemes is less than 0.1\%.

\subsection{Statistics of Korean}
\label{app: ko}

\begin{figure}[ht]
\centering
\includegraphics[width=\linewidth]{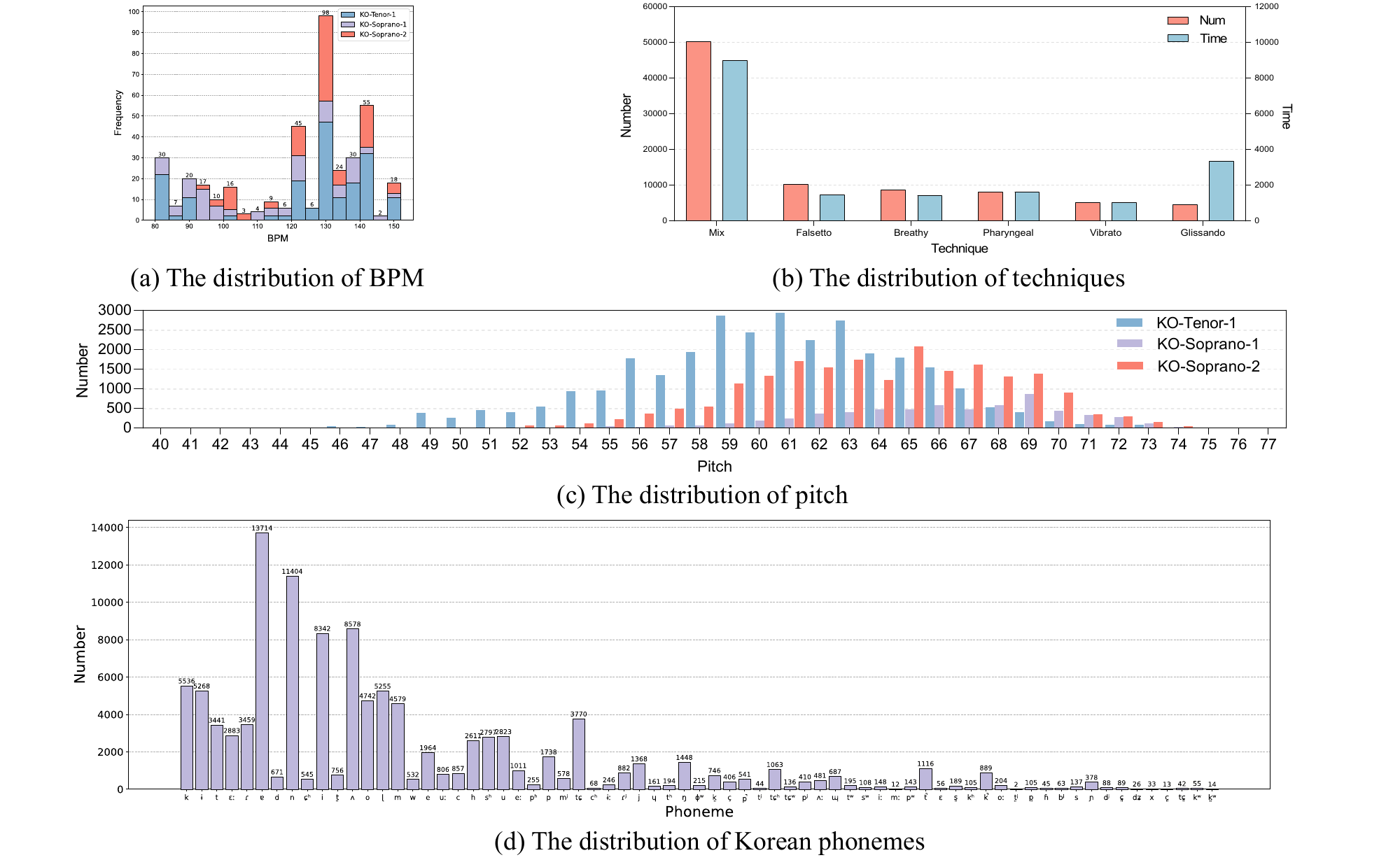}
\vskip-0.1em
\caption{
The statistical distribution of the BPM, techniques, pitch, and phonemes in Korean.
}
\label{fig: ko}
\end{figure}

There are three Korean singers, namely KO-Tenor-1, KO-Soprano-1, and KO-Soprano-2. 
As shown in Figure \ref{fig: ko} (a), the overall BPM distribution ranges from 80 to 150, with a dense distribution between 110 and 150 due to the frequent use of their popular singing styles, which are typically faster.
As illustrated in Figure \ref{fig: ko} (b), the three singers performed a variety of six techniques, with mixed voice being the most prevalent, glissando being the least used technique in number, and pharyngeal being the least used technique in time.
Figure \ref{fig: ko} (c) shows that the overall note pitch ranges of realistic music scores span from MIDI note number 55 (G3, 196 Hz) to 69 (A4, 440 Hz). 
KO-Tenor-1 primarily ranges from 53 (F3, 174.61 Hz) to 68 (G\#4, 415.30 Hz), peaking at 61 (C\#4, 277.18 Hz). 
KO-Soprano-1 primarily ranges from 61 (C\#4, 277.18 Hz) to 72 (C5, 523.25 Hz), peaking at 69 (A4, 440 Hz). 
KO-Soprano-2 primarily ranges from 57 (A3, 220 Hz) to 70 (A\#4, 466.16 Hz), peaking at 65 (F4, 349.23 Hz).
Since the popular songs in Korean typically have a relatively high range, we choose three singers whose vocal range is high. 
And this distribution aligns with their vocal ranges.
As mentioned above, Korean phonemes are entirely annotated according to the MFA standard.
Figure \ref{fig: ko} (d) illustrates that the most common phoneme is "\textipa{\textturna}", with 14,393 occurrences.
Like Japanese, phonemes with an occurrence of less than 10 are rarely found in Korean songs, with the total usage frequency of all words containing these phonemes in daily life less than 0.1\%.

\subsection{Statistics of Russian}
\label{app: ru}

\begin{figure}[ht]
\centering
\includegraphics[width=\linewidth]{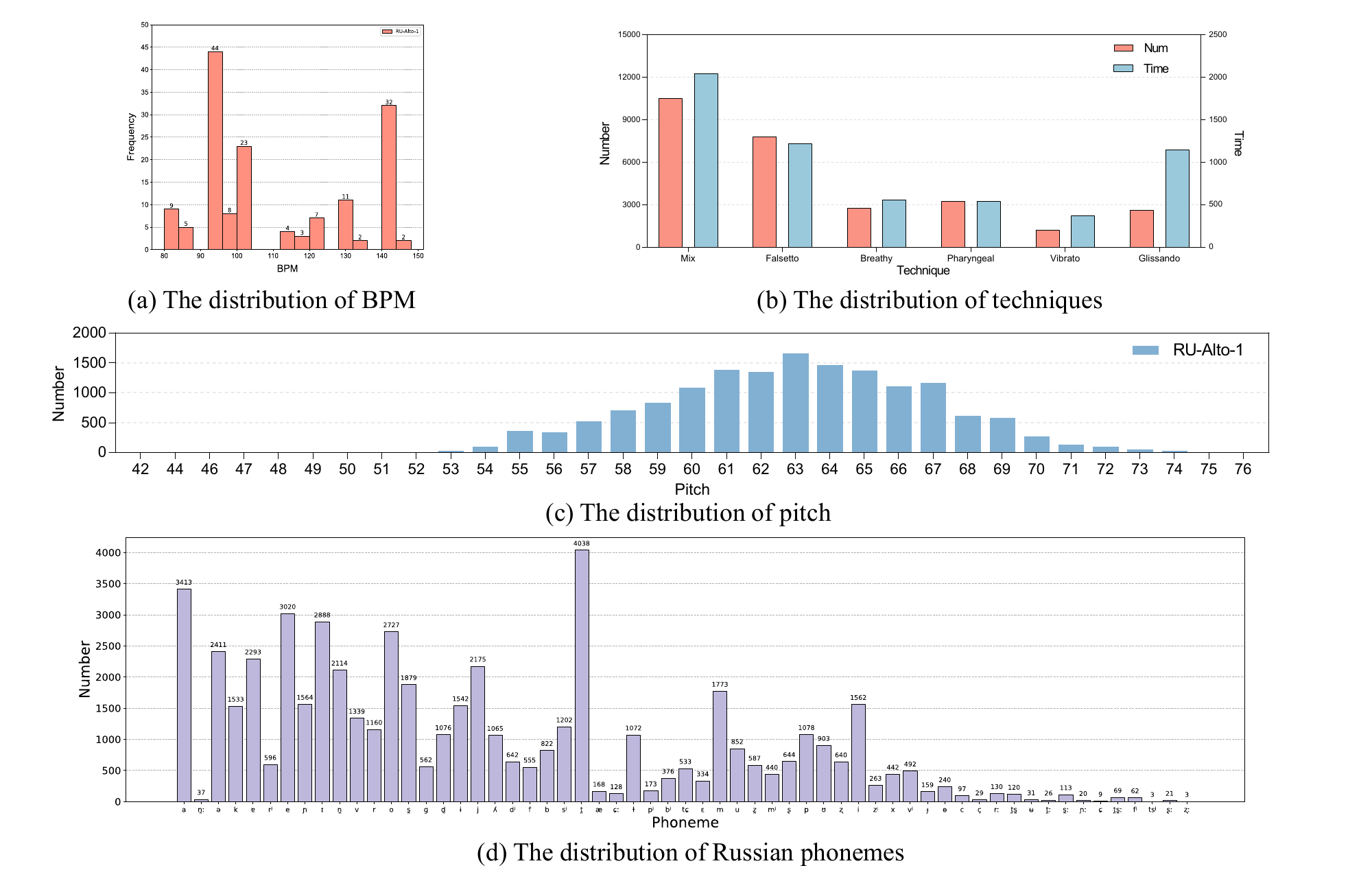}
\vskip-0.1em
\caption{
The statistical distribution of the BPM, techniques, pitch, and phonemes in Russian.
}
\label{fig: ru}
\end{figure}

There is one Russian singer, namely RU-Alto-1.
As shown in Figure \ref{fig: ru} (a), the overall BPM distribution ranges from 80 to 150.
As illustrated in Figure \ref{fig: ru} (b), the singer performs a variety of six techniques. Like the majority of technique distribution, mixed voice is the most prevalent and vibrato is the least used technique.
Figure \ref{fig: ru} (c) shows that the overall note pitch ranges of realistic music scores span from MIDI note number 57 (A3, 220 Hz) to 69 (A4, 440 Hz), peaking at 63 (D\#4, 311.13 Hz).
This distribution aligns with the singer's vocal range.
As mentioned above, Russian phonemes are entirely annotated according to the MFA standard and shown in Figure \ref{fig: ru} (d). 
Like other languages, the total usage frequency of all words containing phonemes with an occurrence of less than 10 in daily life is less than 0.1\%.

\subsection{Statistics of Spanish}
\label{app: es}

\begin{figure}[ht]
\centering
\includegraphics[width=\linewidth]{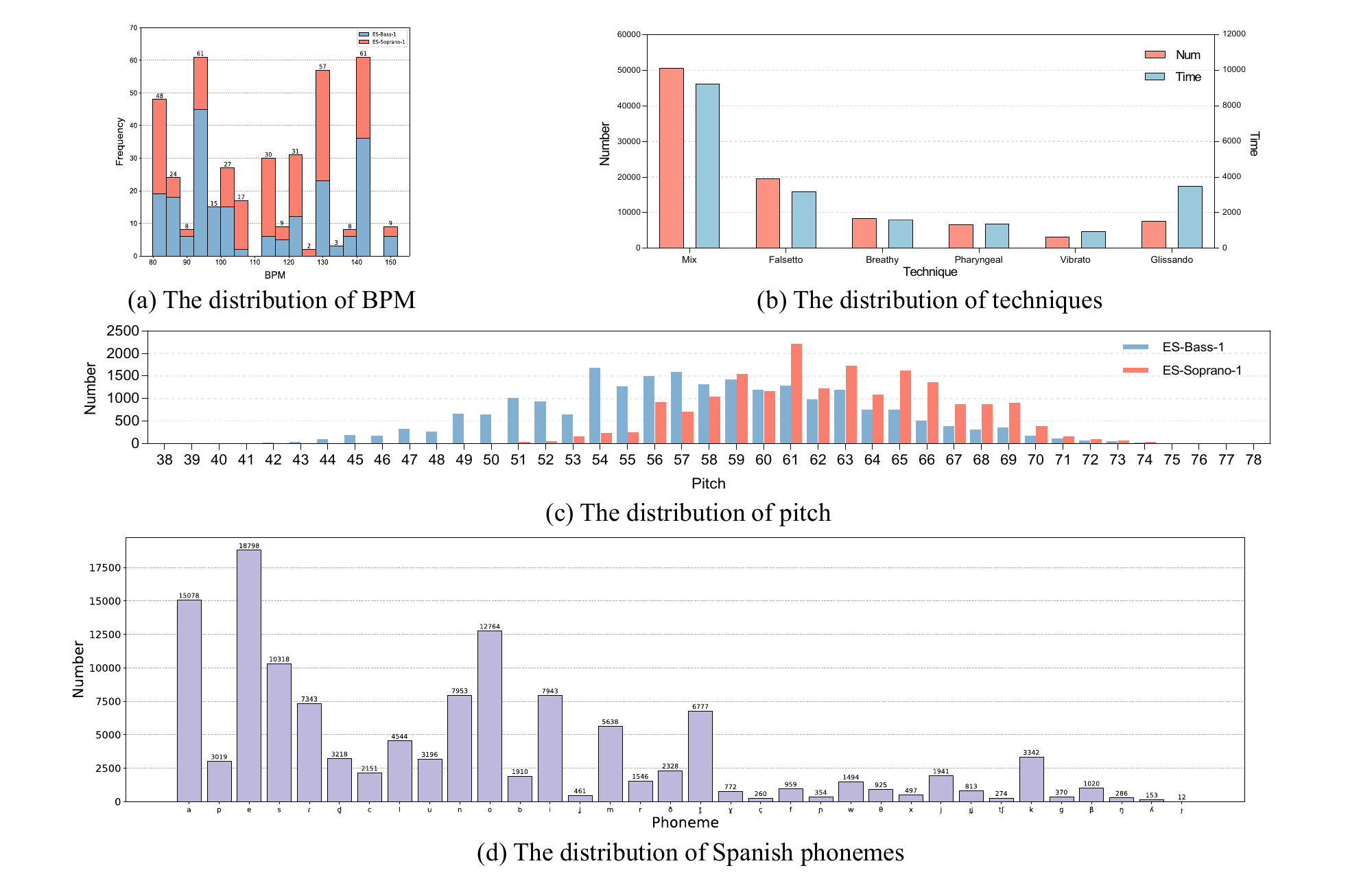}
\vskip-0.1em
\caption{
The statistical distribution of the BPM, techniques, pitch, and phonemes in Spanish.
}
\label{fig: es}
\end{figure}

There are two Spanish singers, namely ES-Bass-1 and ES-Soprano-1. 
As shown in Figure \ref{fig: es} (a), the overall BPM distribution ranges from 80 to 150. As illustrated in Figure \ref{fig: es} (b), the two singers perform a variety of six techniques, with mixed voice being the most prevalent and vibrato being the least used technique. 
Figure \ref{fig: es} (c) shows that the overall note pitch ranges of realistic music scores span from MIDI note number 49 (C\#3, 138.59 Hz) to 69 (A4, 440 Hz). 
ES-Bass-1 primarily ranges from 49 (C\#3, 138.59 Hz) to 65 (F4, 349.23 Hz), peaking at 54 (F\#3, 185 Hz). 
ES-Soprano-1 primarily ranges from 56 (G\#3, 207.65 Hz) to 69 (A4, 440 Hz), peaking at 61 (C\#4, 277.18 Hz).
This distribution aligns with their vocal ranges. 
As mentioned above, Spanish phonemes are entirely annotated according to the MFA standard.
Figure \ref{fig: es} (d) illustrates that the most common phoneme is "e", with 18,798 occurrences, while the least common phonemes are "\textipa{\textbardotlessj}", with 12 occurrences. 
The missing phoneme is "\textipa{S}", which is barely seen in Spanish songs, and the sum usage frequency of all words containing it in daily life is less than 0.1\%.

\subsection{Statistics of French}
\label{app: fr}

\begin{figure}[ht]
\centering
\includegraphics[width=\linewidth]{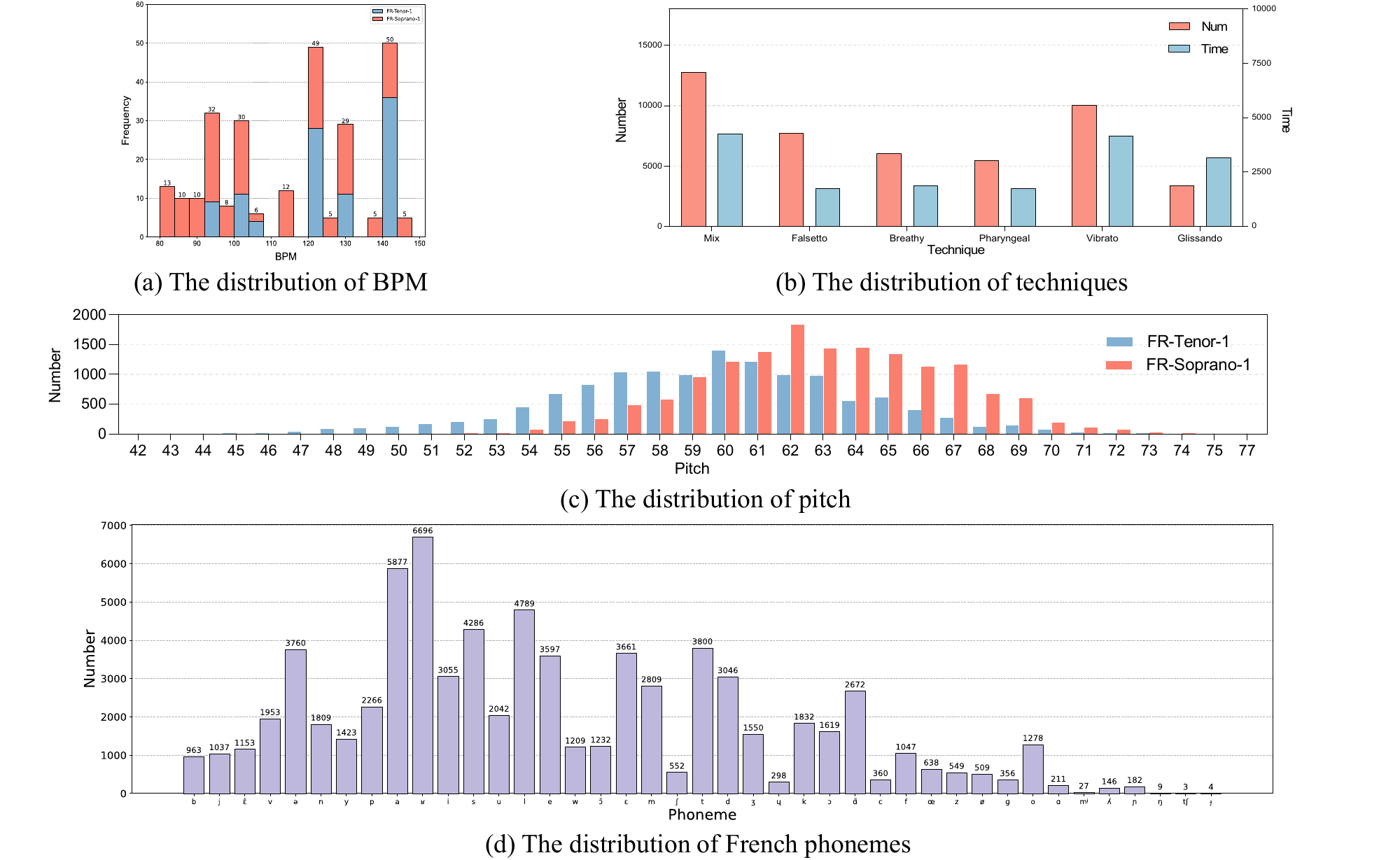}
\vskip-0.1em
\caption{
The statistical distribution of the BPM, techniques, pitch, and phonemes in French.
}
\label{fig: fr}
\end{figure}

There are two French singers, namely FR-Tenor-1 and FR-Soprano-1. 
As shown in Figure \ref{fig: fr} (a), the overall BPM distribution ranges from 80 to 150. 
As illustrated in Figure \ref{fig: fr} (b), the two singers perform a variety of six techniques, with mixed voice being the most prevalent, pharyngeal being the least used technique in number, and glissando being the least used technique in time. 
Figure \ref{fig: fr} (c) shows that the overall note pitch ranges of realistic music scores span from MIDI note number 54 (F\#3, 185 Hz) to 69 (A4, 440 Hz). 
FR-Tenor-1 primarily ranges from 54 (F\#3, 185 Hz) to 65 (F4, 349.23 Hz), peaking at 60 (C4, 261.63 Hz). 
FR-Soprano-1 primarily ranges from 57 (A3, 220 Hz) to 69 (A4, 440 Hz), peaking at 62 (D4, 293.66 Hz).
This distribution aligns with their vocal ranges. 
As mentioned above, French phonemes are entirely annotated according to the MFA standard.
Figure \ref{fig: fr} (d) illustrates that the most common phoneme is "
\textipa{\textinvscr}", with 6,696 occurrences, while the least common phonemes are "\textipa{tS}" with 3 occurrences.
Like other languages, the total usage frequency of all words containing phonemes with an occurrence of less than 10 in daily life is less than 0.1\%.

\subsection{Statistics of German}
\label{app: de}

\begin{figure}[ht]
\centering
\includegraphics[width=\linewidth]{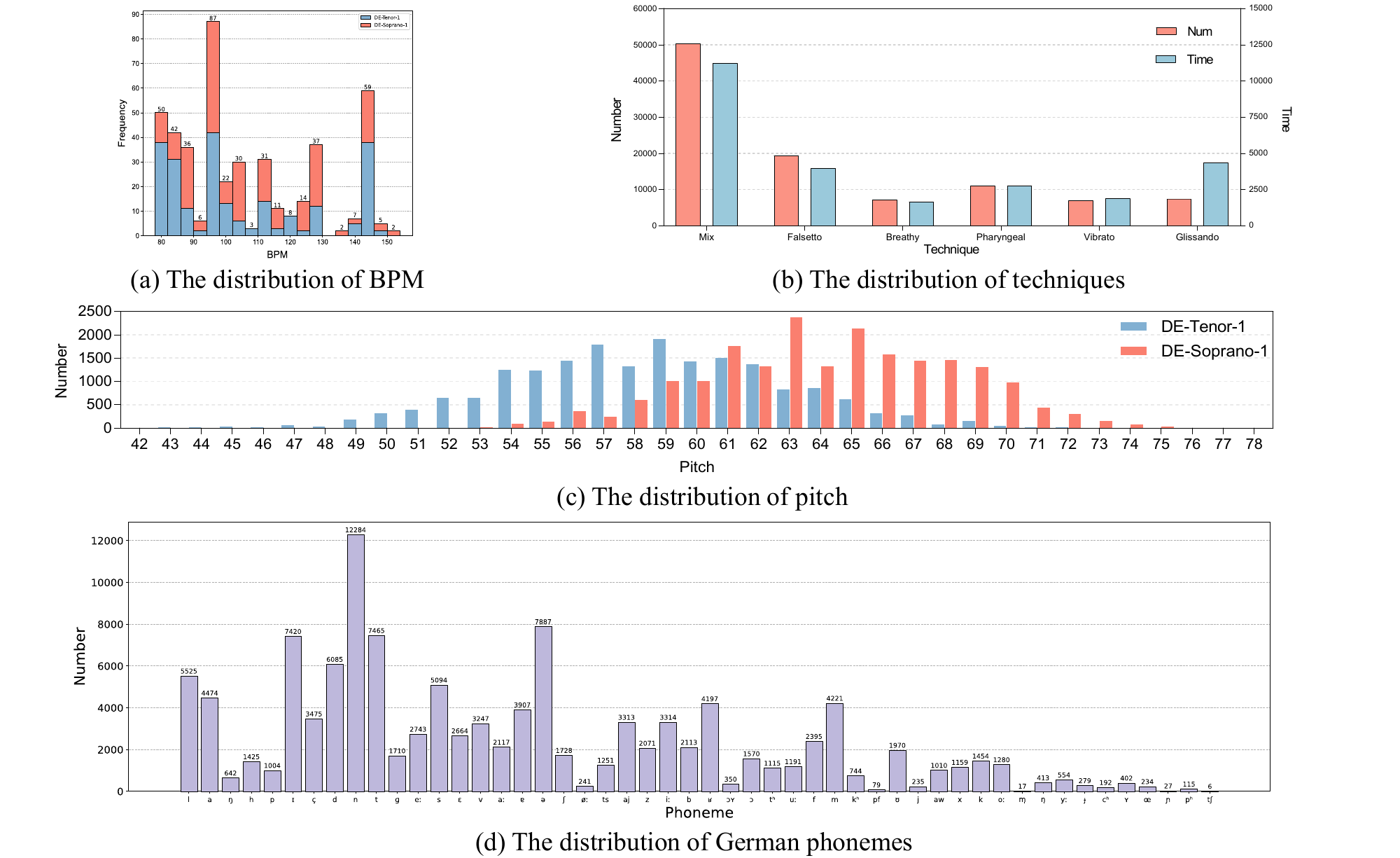}
\vskip-0.1em
\caption{
The statistical distribution of the BPM, techniques, pitch, and phonemes in German.
}
\label{fig: de}
\end{figure}

There are two German singers, namely DE-Tenor-1 and DE-Soprane-1. 
As shown in Figure \ref{fig: de} (a), the overall BPM distribution ranges from 80 to 150, which has a dense distribution between 80 and 110 due to the bel canto singing method for the singing, which is typically slow.
As illustrated in Figure \ref{fig: de} (b), the three singers perform a variety of six techniques, with mixed voice being the most prevalent and vibrato being the least used technique.
Figure \ref{fig: de} (c) shows that the overall note pitch ranges of realistic music scores span from MIDI note number 52 (E3, 164.81 Hz) to 71 (B4, 493.88 Hz). 
DE-Tenor-1 primarily ranges from 52 (E3, 164.81 Hz) to 65 (F4, 349.23 Hz), peaking at 59 (B3, 246.94 Hz). 
DE-Soprano-1 primarily ranges from 58 (A\#3, 233.08 Hz) to 71 (B4, 493.88 Hz), peaking at 63 (D\#4, 311.13 Hz).
This distribution aligns with their vocal ranges.
As mentioned above, German phonemes are entirely annotated according to the MFA standard.
Figure \ref{fig: de} (d) illustrates that the most common phoneme is "n", with 12,284 occurrences,  while the least common is "\textipa{tS}", with 6 occurrences. 
Like other languages, the total usage frequency of all words containing phonemes with an occurrence of less than 10 in daily life is less than 0.1\%.

\subsection{Statistics of Italian}
\label{app: it}

\begin{figure}[ht]
\centering
\includegraphics[width=\linewidth]{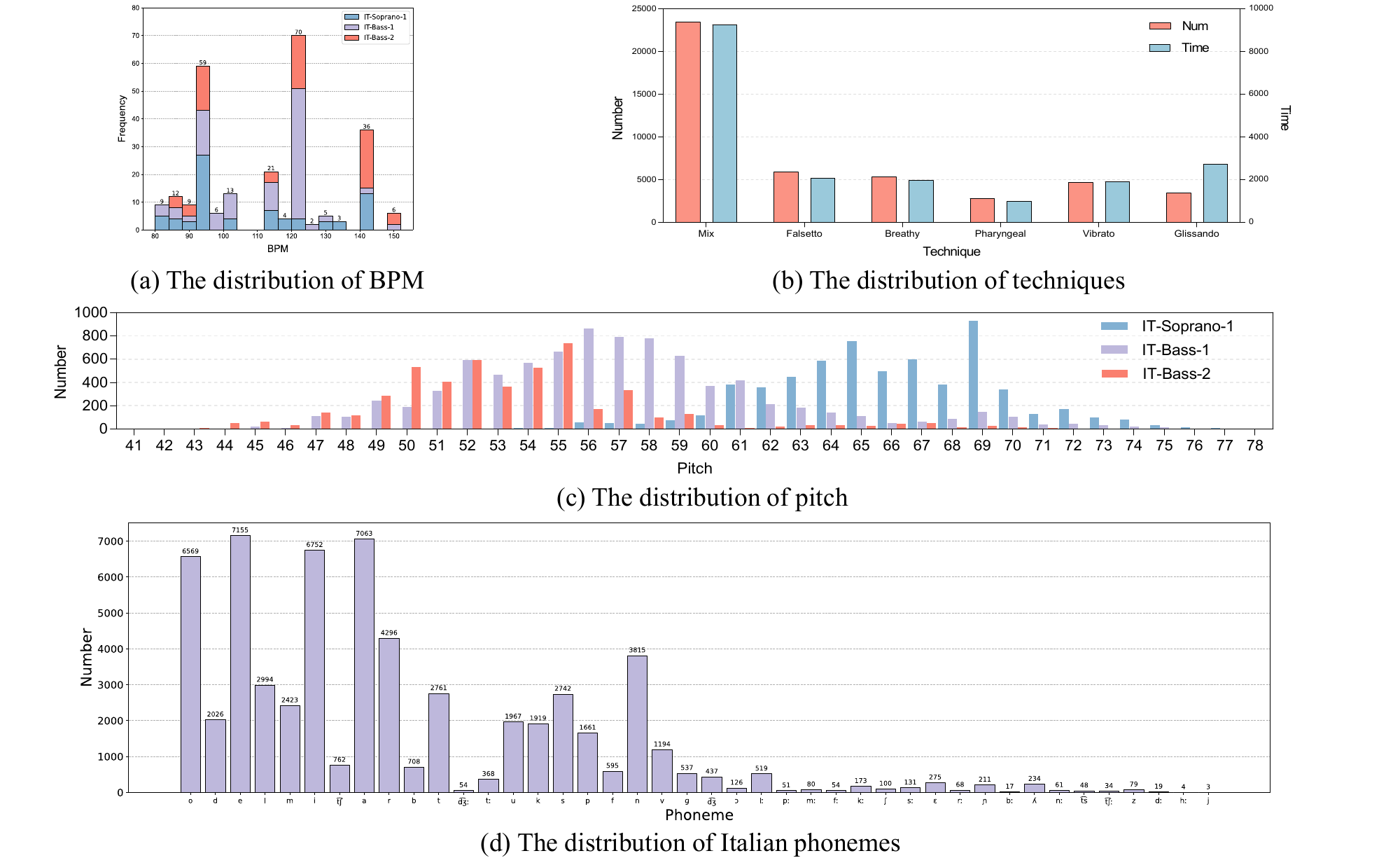}
\vskip-0.1em
\caption{
The statistical distribution of the BPM, techniques, pitch, and phonemes in Italian.
}
\label{fig: it}
\end{figure}

There are three Italian singers, namely IT-Bass-1, IT-Bass-2, and IT-Soprano-1. 
As shown in Figure \ref{fig: it} (a), the overall BPM distribution ranges from 80 to 150.
As illustrated in Figure \ref{fig: it} (b), the three singers perform a variety of six techniques, with mixed voice being the most prevalent and pharyngeal being the least used technique.
Figure \ref{fig: it} (c) shows that the overall note pitch ranges of realistic music scores span from MIDI note number 49 (C\#3, 138.59 Hz) to 70 (A\#4, 466.16 Hz). 
IT-Bass-1 primarily ranges from 61 (C\#4, 277.18 Hz) to 71 (B4, 493.88 Hz), peaking at 69 (A4, 440 Hz). 
IT-Bass-2 primarily ranges from 49 (C\#3, 138.59 Hz) to 63 (D\#4, 311.13 Hz), peaking at 56 (G\#3, 207.65 Hz). 
IT-Soprano-1 primarily ranges from 49 (C\#3, 138.59 Hz) to 59 (B3, 246.94 Hz), peaking at 55 (G3, 196 Hz).
This distribution aligns with their vocal ranges.
As mentioned above, Italian phonemes are entirely annotated according to the Epitran standard, which performs best for Italian phoneme alignment.
Figure \ref{fig: it} (d) illustrates that the most common phoneme is "e", with 7,155 occurrences, while the least common phonemes are "j" and "\textipa{h\textlengthmark}", with 3 and 4 occurrences, respectively.
Like other languages, the total usage frequency of all words containing phonemes with an occurrence of less than 10 in daily life is less than 0.1\%.

\section{Details of Experiments}
\label{app: exp}

\subsection{Subjective Evaluation}
\label{app: sub}

For each task, we randomly select 50 sentences from our test set for subjective evaluation, each of which is listened to by at least 20 professional listeners.
To evaluate the model performance, we conduct the MOS (Mean opinion score) evaluation.
In the context of MOS-Q evaluations, these listeners are instructed to concentrate on synthesis quality (including clarity, naturalness, and rich stylistic details), irrespective of singer similarity (in terms of timbre and styles). 
Conversely, during MOS-S evaluations, the listeners are directed to assess singer similarity (singer similarity in terms of timbre and styles) to the audio reference, disregarding any differences in content or synthesis quality (including quality, clarity, naturalness, and rich stylistic details). 
For MOS-C, the listeners are informed to evaluate technique controllability (accuracy and expressiveness of technique control), disregarding any differences in content, timbre, or synthesis quality (including quality, clarity, naturalness, and rich stylistic details). 
In both MOS-Q, MOS-S, and MOS-C evaluations, listeners are requested to grade various singing voice samples on a Likert scale ranging from 1 to 5. 

\begin{figure}[ht]
\centering
\includegraphics[width=\linewidth]{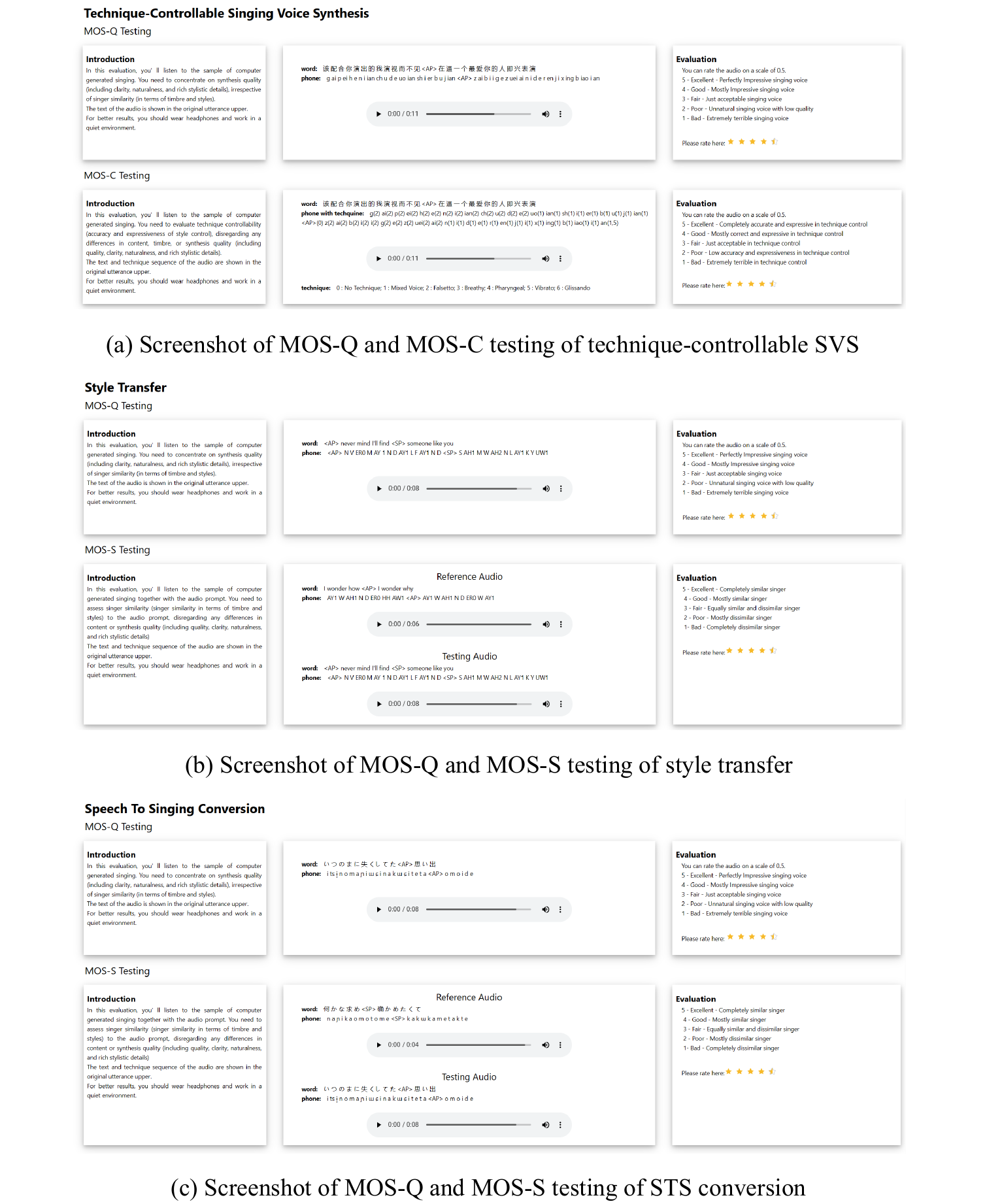}
\vskip-0.1em
\caption{
The statistical distribution of screenshots of MOS testings for diverse tasks.
}
\label{fig: sub}
\end{figure}

The screenshots of instructions for testers are shown in Figure \ref{fig: sub}. 
It is important to note that all participants are fairly compensated for their time and effort. 
We compensated participants at a rate of \$15 per hour, resulting in a total expenditure of approximately \$2000 on participant compensation.
Participants are informed that the results will be used for scientific research.

\subsection{Objective Evaluation}
\label{app: obj}

For our evaluation, we select various objective metrics tailored to different tasks.
First, we use F0 Frame Error (FFE), which combines voicing decision error and F0 error metrics to capture F0 information comprehensively.
Next, we employ Mean Cepstral Distortion (MCD) for measuring audio quality as the formula:
\begin{equation}
\begin{aligned}
&\text{MCD} = \frac{10}{\ln 10} \sqrt{2 \sum_{d=1}^{D} (c_t(d) - \hat{c}_t(d))^2},
\end{aligned}
\end{equation}
where \(c_t(d)\) and \(\hat{c}_t(d)\) represent the \(d\)-th MFCC of the target and predicted frames at time \(t\), respectively, and \(D\) is the number of MFCC dimensions.
Additionally, Cosine Similarity (Cos) is utilized to quantify the resemblance between the synthesized and reference singing voices. We calculate the average cosine similarity between the embeddings extracted from the synthesized voices and the ground truth.
We use the WavLM \cite{chen2022wavlm} fine-tuned for speaker verification \footnote{https://huggingface.co/microsoft/wavlm-base-plus-sv} to extract singer embedding.

\subsection{Technique-Controllable Singing Voice Synthesis}
\label{app: tc}

For each model, we incorporated a simple technique embedding. 
This module takes as input six binary sequences indicating the presence or absence of each technique on each phoneme. 
These sequences are then encoded and concatenated.
In the non-parallel experiments, we randomly generate technique sequences and select the appropriate ones to input into the model, assigning zero, one, or multiple techniques to each target phoneme.

\begin{figure}[ht]
\centering
\includegraphics[width=\linewidth]{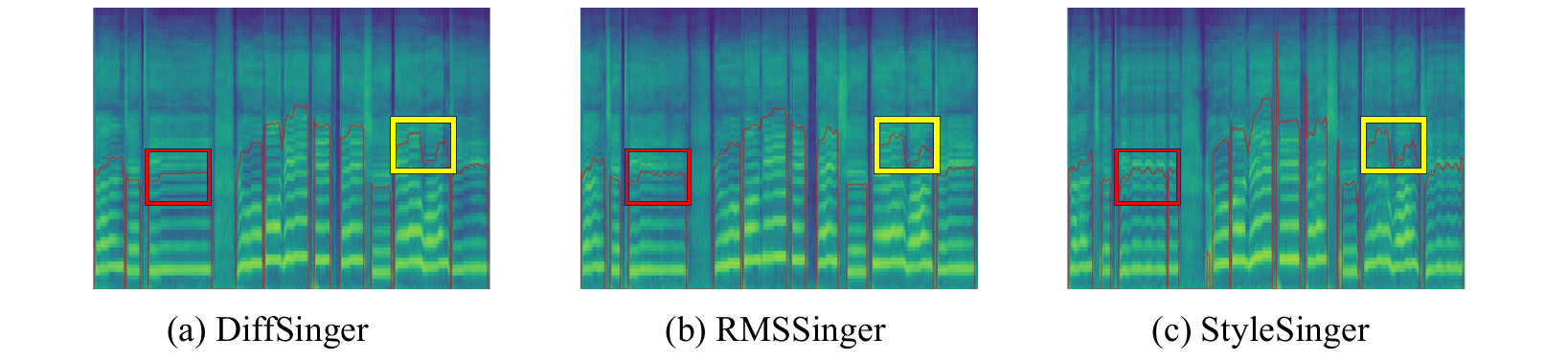}
\vskip-0.1em
\caption{
The mel-spectrograms depict the results of technique-controllable SVS. 
The vibrato technique is indicated by red boxes, and yellow boxes show the mixed voice technique.
}
\label{fig: svs}
\end{figure}

We used a sequence incorporating mixed voice and vibrato techniques for technique control. 
As shown in Figure \ref{fig: svs}, DiffSinger failed to control vibrato, whereas StyleSinger enhanced the expressiveness of mixed voice through detailed styles.

\subsection{Technique Recognition}
\label{app: tr}

We design a technique recognition model based on ROSVOT \cite{li2024robust}.
This model is originally designed for coarse notes of music score detection.
Simply, we change the loss to the cross entropy loss between the GT and predicted technique labels.
The inputs of the technique recognition model include mel-spectrogram, pitch, and phoneme boundaries, with the output being the predicted probabilities of six distinct techniques.

\begin{table}[ht]
\centering
\small
\caption{
Precision, Recall, F1, and Accuracy of each technique in both overall and cross-lingual technique recognition. 
We provide both Asian-to-European and European-to-Asian results.
}
\scalebox{0.98}{
\begin{tabular}{l|c|cccccc}
\toprule
\multirow{2}{*}{\bfseries{Experiment}} & \multirow{2}{*}{\bfseries{Metric}} & \multicolumn{6}{c}{\textbf{Technique Recognition Accuracy}}\\
& &{mixed voice} & {falsetto} &{breathy} & {pharyngeal} &{vibrato} & {glissando}\\
\midrule
\multirow{4}{*}{\textbf{Overall}} & Precision & 0.95 & 0.98 & 0.99 & 0.96 & 0.75 & 0.75 \\
& Recall & 0.71 & 0.95 & 0.99 & 0.82 & 0.72 & 0.71 \\
& F1 & 0.78 & 0.96 & 0.99 & 0.85 & 0.70 & 0.70 \\
& Accuracy & 0.78 & 0.84 & 0.78 & 0.80 & 0.89 & 0.85 \\
\midrule
\multirow{4}{*}{\textbf{Asian-to-European}} & Precision & 0.92 & 0.97 & 0.99 & 0.97 & 0.62 & 0.63 \\
& Recall & 0.55 & 0.93 & 0.98 & 0.79 & 0.68 & 0.63 \\
& F1 & 0.65 & 0.94 & 0.98 & 0.84 & 0.58 &  0.51 \\
& Accuracy & 0.74 & 0.78 & 0.71 & 0.78 & 0.87 & 0.76 \\
\midrule
\multirow{4}{*}{\textbf{European-to-Asian}} & Precision & 0.89 & 0.95 & 0.93 & 0.87 & 0.72 & 0.61 \\
& Recall & 0.64 & 0.90 & 0.95 & 0.77 & 0.61 & 0.68 \\
& F1 & 0.71 & 0.92 & 0.93 & 0.79 & 0.57 & 0.58 \\
& Accuracy & 0.76 & 0.8 & 0.73 & 0.76 & 0.81 & 0.81 \\
\bottomrule
\end{tabular}}
\label{tab: tr2}
\end{table}

As shown in Table \ref{tab: tr2}, our model achieves high Precision, Recall, F1, and Accuracy across all six techniques, both in overall and cross-lingual experiments.
This performance also underscores the quality of controlled comparison and phoneme-level technique annotations in GTSinger.

\subsection{Style Transfer}
\label{app: st}

\begin{figure}[ht]
\centering
\includegraphics[width=\linewidth]{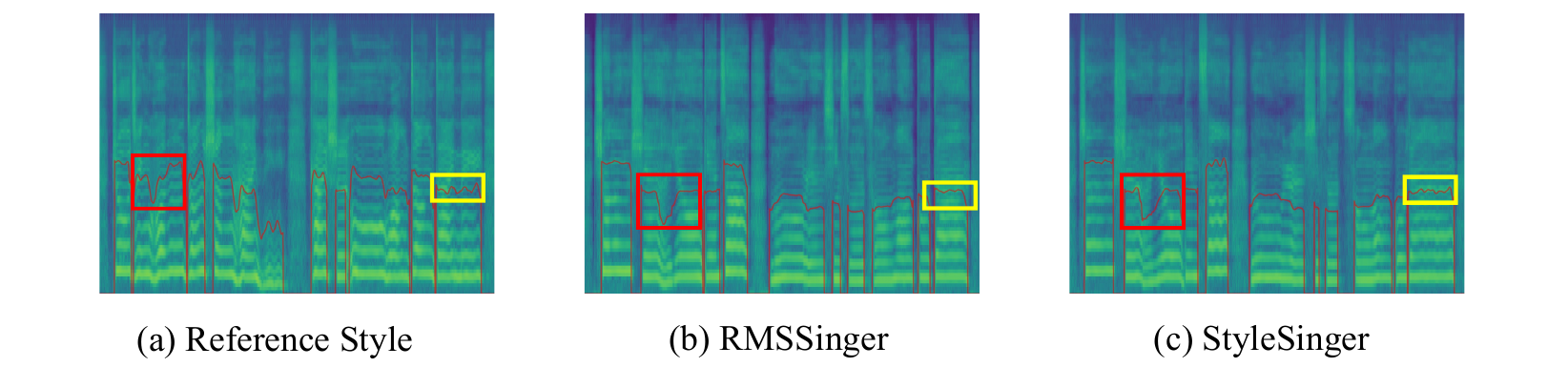}
\vskip-0.1em
\caption{
The mel-spectrograms depict the results of style transfer. 
The vibrato style is indicated by yellow boxes, and the pronunciation and articulation skills are highlighted in red boxes.
}
\label{fig: st}
\end{figure}

We proceed to visualize mel-spectrograms and pitch contours of style transfer experiments in Figure \ref{fig: st}. 
StyleSinger excels at capturing the intricate nuances of the reference style. 
The pitch curve generated by StyleSinger exhibits a greater range of variations and finer details, effectively capturing the vibrato technique, as well as the nuances of pronunciation and articulation skills. 
In contrast, the curves generated by RMSSinger appear relatively flat.
Additionally, StyleSinger excels in modeling mel-spectrograms with higher quality and stylistic details. 

\subsection{Speech-to-Singing Conversion}
\label{app: sts}

We modify StyleSinger to use the paired speech reference input for generating the singing voice, enabling it to perform STS conversion and fully model the style transfer between speech and singing voice. 
While AlignSTS requires the input of the singing voice's f0 for conversion, StyleSinger only needs realistic music scores, making it more suitable for practical applications.

\begin{figure}[ht]
\centering
\includegraphics[width=\linewidth]{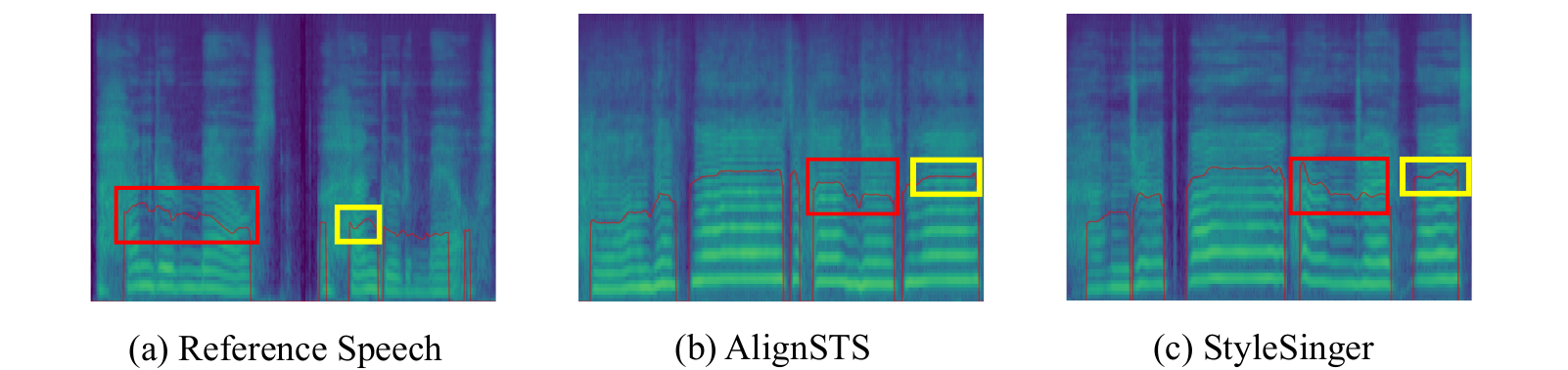}
\vskip-0.1em
\caption{
The mel-spectrograms depict the results of STS conversion. 
The pronunciation and articulation skills are highlighted in yellow boxes, and the pitch transitions are shown in red boxes.
}
\label{fig: sts}
\end{figure}

We visualize mel-spectrograms and pitch contours of STS experiments in Figure \ref{fig: sts}. 
We observe that StyleSinger successfully transfers the pronunciation and articulation skills, as well as the pitch transition styles. 
Compared to AlignSTS, which exhibits flat pitch with a lack of style variation and mel-spectrograms lacking in detail, StyleSinger demonstrates more pitch variations that closely match the reference speech style, along with higher quality mel-spectrograms.

\end{document}